\documentclass[10.5pt,nofootinbib,showpacs]{revtex4-2}
\usepackage{amsmath,graphicx}
\usepackage{dcolumn}
\usepackage{bm}
\usepackage{color}
\usepackage{epstopdf}
\usepackage{eurosym}
\usepackage{multirow}
\usepackage{wrapfig}
\usepackage{amsfonts}
\usepackage{amsmath}
\usepackage{hyperref}
\usepackage{graphicx,color}
\usepackage{amssymb}
\usepackage[english]{babel}
\usepackage{graphicx}
\usepackage{epsfig}
\usepackage{bm}
\usepackage{longtable}
\usepackage{verbatim}
\usepackage{longtable}
\usepackage{color}
\usepackage{subfigure}
\usepackage{multirow}
\usepackage{rotating}
\usepackage{makecell}
\usepackage{pbox}
\usepackage{textcomp}%
\usepackage{booktabs}

\begin{document}
\newcommand{\hs}{\hspace*{0.25cm}}
\newcommand{\ms}{\hspace*{1cm}}
\newcommand{\vs}{\vspace*{0.5cm}}
\newcommand{\be}{\begin{equation}}
\newcommand{\ee}{\end{equation}}
\newcommand{\bea}{\begin{eqnarray}}
\newcommand{\eea}{\end{eqnarray}}
\newcommand{\ben}{\begin{enumerate}}
\newcommand{\een}{\end{enumerate}}
\newcommand{\bde}{\begin{widetext}}
\newcommand{\ede}{\end{widetext}}
\newcommand{\nn}{\nonumber}
\newcommand{\crn}{\nonumber \\}
\newcommand{\Tr}{\mathrm{Tr}}
\newcommand{\non}{\nonumber}
\newcommand{\noi}{\noindent}
\newcommand{\al}{\alpha}
\newcommand{\la}{\lambda}
\newcommand{\bet}{\beta}
\newcommand{\ga}{\gamma}
\newcommand{\va}{\varphi}
\newcommand{\om}{\omega}
\newcommand{\pa}{\partial}
\newcommand{\+}{\dagger}
\newcommand{\fr}{\frac}
\newcommand{\sq}{\sqrt}
\newcommand{\bc}{\begin{center}}
\newcommand{\ec}{\end{center}}
\newcommand{\Ga}{\Gamma}
\newcommand{\de}{\delta}
\newcommand{\De}{\Delta}
\newcommand{\ep}{\epsilon}
\newcommand{\varep}{\varepsilon}
\newcommand{\ka}{\kappa}
\newcommand{\La}{\Lambda}
\newcommand{\si}{\sigma}
\newcommand{\Si}{\Sigma}
\newcommand{\ta}{\tau}
\newcommand{\up}{\upsilon}
\newcommand{\Up}{\Upsilon}
\newcommand{\ze}{\zeta}
\newcommand{\ps}{\psi}
\newcommand{\Ps}{\Psi}
\newcommand{\ph}{\phi}
\newcommand{\vph}{\varphi}
\newcommand{\Ph}{\Phi}
\newcommand{\Om}{\Omega}
\newcommand{\Blue}[1]{{\color{blue}#1}}
\newcommand{\Red}[1]{{\color{red}#1}}
\newcommand{\Revised}[1]{{\color{red}#1}}
\title{Neutrino mass and mixing, resonant leptogenesis and charged lepton flavor violation in a minimal inverse seesaw model with $S_4$ symmetry}
\author{V. V. Vien$^{a}$}
\email{vvvien@ttn.edu.vn}
\affiliation{$^{a}$Department of Physics, Tay Nguyen University, 
Daklak province, Vietnam.}
\author{Mayengbam Kishan Singh$^{b}$}
\email{kishanmayengbam@gmail.com}
\affiliation{$^{b}$Department of Physics, D.M. College of Science, Dhanamanjuri University, Manipur 795001, 
India.}

\date{\today}
\begin{abstract}
We propose a minimal inverse seesaw model with $S_4$ symmetry for the Majorana neutrinos with only one real ($m_0$)-and two complex ($\alpha, \beta$) parameters in neutrino sector which gives reasonable predictions for the neutrino oscillation parameters, the observed baryon asymmetry of the Universe and the charged lepton flavor violation. The resulting model reveals a favor for normal neutrino mass ordering, a higher octant of $\theta_{23}$ and a lower half-plane of Dirac CP violation phase. The predictions of the model for sum of neutrino masses and the effective Majorana neutrino mass are centered around 58.98 meV and 6.2 meV, respectively. The model also provides the predictions of the baryon asymmetry and charged lepton flavour violation processes which are consistent with the experimental observations.
\end{abstract}

\maketitle
\section{Introduction}
Despite its great success, the Standard Model (SM) has significant limitations. Some of the notable limitations of the SM is derived from  neutrino oscillation data, the observed baryon asymmetry of the
Universe (BAU) and the charged lepton flavor violation (cLFV).

The neutrino oscillation parameters have now been observed with certain precision by many experiments. However, a global analysis will yield results more consistent
than an individual one. Various studies have carried out
the global analysis of neutrino oscillation data such as Refs. \cite{Salas2021,Esteban2020,Capozzi2021}. We use the data in Ref.\cite{Salas2021} (see Table \ref{Salas2021T}) for the numerical analysis.

\begin{table}[ht]
\begin{center}
\vspace{-0.15 cm}
\caption{The neutrino oscillation data for normal hierarchy (NH) \cite{Salas2021}} \label{Salas2021T}
\begin{tabular}{@{}llll@{}} \toprule
 Parameter&3$\sigma$  (best-fit point) \\ \midrule 
$s^{2}_{12}/10^{-1}$&$2.71\rightarrow3.69 \,(3.18)$ \\ 
$s^{2}_{13}/10^{-2}$&$2.000\rightarrow2.405\, (2.200)$\\ 
$s^{2}_{23}/10^{-1}$&$4.34\rightarrow 6.10 \,(5.74)$\\ 
$\delta/\pi$&$0.71\rightarrow 1.99 \,(1.08)$\\ 
$| \Delta m^{2}_{31}|[10^{-3}\, \mathrm{eV}^2]$\hspace*{0.2 cm}&$2.47\rightarrow2.63\, (2.55)$\\ 
$\Delta m^{2}_{21}[10^{-5}\, \mathrm{eV}^2]$\hspace*{0.2 cm}&$6.94\rightarrow 8.14\, (7.50)$\\ \toprule 
\end{tabular}
\end{center}
\end{table}

On the other hand, the BAU, defined by the baryon-to-photon density ratio, $\eta_B = \big(n_B - n_{\bar{B}}\big)/n_\gamma$,
where $n_B (n_{\bar{B}})$ and $n_\gamma$ respectively denotes the number densities of baryons (antibaryons) and photons.
The value of BAU
can be deduced from observations via big bang nucleosynthesis \cite{Cyburt2005,Yao2006,Steigman2006}.
The constraint on BAU is given by \cite{NAghanim2020},
\bea
6.08\leq 10^{10}\eta_B\leq 6.16. \label{baryonbound}
\eea
In the SM, the cLFV processes are highly suppressed \cite{Calibbi2017uvl,Ardu2022sbt,Cei2014jtm,Matsuzaki2008ik,CelisPRD14,Omura2015xcg,Zhou2016ynv,PhysRevD.99.035020,HernandezTome2018fbq,
Calcuttawala2018wgo,Ferreira2019qpf,Baldiniẹpc16,Afanacievẹpc24}, however, upcoming experiments are improving the sensitivities to search for these  processes. Hence, cLFV is also one of the valuable indicators
of physics beyond the SM. Since muons own a longer lifetime than tauons
and are copiously generated
in cosmic radiation, transitions involving muons become the most interesting issue. In this work, we will consider
the decay $\mu \rightarrow \gamma e$ whose limit on the branching ratio has been reported by
MEG II \cite{Baldiniẹpc16,MegII2024data}, with
\begin{equation}
\text{BR}(\mu \rightarrow \gamma e) < 3.1 \times 10^{-13}.
\end{equation}
By the end of 2026, MEG II targets a sensitivity of $\sim 6 \times 10^{-14}$ on BR$(\mu \rightarrow e\gamma)$ \cite{Afanacievẹpc24}.
The current Belle limits on BR$(\tau \rightarrow e\gamma)$ \cite{Aubert2010} and BR$(\tau \rightarrow \mu\gamma)$ \cite{Abdesselam21} are $3.3\times 10^{-8}$
and $4.2 \times 10^{-8}$,
respectively. Belle II is expected to reach a limit of $\mathcal{O}(10^{-9})$ \cite{Kouptep19}.

The neutrino mass matrix ($m_\nu$), which can be originated from the Yukawa-like couplings and can be generated from the see-saw mechanisms, is an important object to understand the neutrino physics. Neutrino and charged-lepton mass ($M_l$) matrices contain information of twelve parameters, including three charged-lepton masses, three neutrino masses, three mixing angles, one Dirac - and two Majorana phases. Specific structures of $m_\nu$ can be generated by the extension of the SM with discrete symmetries in the light of the seesaw mechanisms \cite{MinkowskiSSI,YanagidaSSI,MannSSI,MohapatraSSI,VienSSI1,VienSSI2,VienSSI3,Mksingh1,Mksingh2,SchechterSSII,SSII,MohapatraSSII,FootSSIII,VienL15,Vien22,VienL23}. The most popular mechanism for generating neutrino masses is the (canonical) seesaw mechanisms \cite{MinkowskiSSI,YanagidaSSI,MannSSI,MohapatraSSI,VienSSI1,VienSSI2,VienSSI3,SchechterSSII,SSII,MohapatraSSII,FootSSIII,VienL15,Vien22,VienL23}, however, they are very difficult to search for heavy particles 
\cite{VissaniHP,CasasHP,GarcíaHP,ArcadiHP}. Hence, low-scale seesaw mechanisms \cite{LindnerLSS,AkhmedovLSS,MalinskyLSS,MohapatraISS1,MohapatraISS2,GarciaISS,DavidsonUS,AntonuioUSS,ChenUSS} become interesting to be investigated. Among low scale schemes, the inverse seesaw mechanism \cite{MohapatraISS1,MohapatraISS2,GarciaISS} is one of the popular ways of producing the small neutrino mass with TeV scale heavy neutrinos, which can be tested by the experiment.

The most feasible and minimal version  of the inverse seesaw mechanism is
 minimal seesaw mechanism ISS(2,2) \cite{AbadaMISS,ThapaJHEP23} in which two singlet neutral fermions and two right-handed neutrinos are added to the SM. Recently, the SM extension with $S_4$ symmetry together with abelian symmetries $Z_3$ and $Z_4$ has been presented in Ref. \cite{ThapaJHEP23} in which up to twelve singlet scalars (flavons) are added to the SM.

It is noted that, in Ref. \cite{ThapaJHEP23}, two SM gauge singlet fermions $S_1$ and $S_2$ are respectively assigned in two different singlets $1_1$ and $1_2$ of $S_4$ symmetry. As a consequence, it is necessary to introduce two $SU(2)_L$ singlet scalars which are respectively assigned in $2$ and $1_1$ of $S_4$ to generate two mass matrices $M_R$ and $\mu$. In this kind of ISS, $M_R$ is generated from the couplings of gauge singlet fermions and right-handed neutrinos whereas $\mu$ is Majorana term related to the coupling $\overline{S}S^c$. Furthermore, $S_4$ group has two
singlets $1_1$ and $1_2$ (where $1_1$ corresponds to a
trivial singlet), one doublet $2$ and two triplets $3_1$ and $3_2$. Therefore, two SM gauge singlet fermions $S_1$ and $S_2$ can be either assigned in two different singlets $1_1$ and $1_2$ or in a doublet $2$ of $S_4$. The first case has been studied in Ref. \cite{ThapaJHEP23} where $M_R$ owns all non-zero elements while $\mu$ is a diagonal matrix. In this study, we consider the second case in which $S_1$ and $S_2$ are assigned in one doublet $2$ of $S_4$. As a consequence, two mass matrices $M_R$ and $\mu$ are generated by only one $SU(2)_L$ singlet scalar $\chi$ put in $1_1$ of $S_4$ which is simpler and completely different from those of Ref. \cite{ThapaJHEP23}.

The remaining part of this study is as follows. Section \ref{model} gives a description of the model. Analytic calculation of neutrino mass and mixing is performed in section \ref{neutrinomixing}. The resonant leptogenesis and cLFV processes are presented in sections \ref{reslepto} and \ref{CLV}, respectively. The numerical analysis is devoted to section \ref{numericalana}. Lastly, some conclusions are drawn in section \ref{conclusion}. Appendix \ref{forbidappen} provides Yukawa terms forbidden by the model symmetries and Appendix \ref{Higgspotential} give a brief description of the scalar potential of the considered model.

\section{\label{model}The model}
We propose a SM extension with $S_4$ symmetry augmented by Abelian symmetries $Z_5, Z_3$ and $Z_2$  to obtain the desired structures for the lepton mass matrices. Simultaneously, two right-handed neutrinos ($\nu_R$) and two singlet neutral leptons ($S$) together with singlet scalars $\varphi_l, \phi_l, \varphi_\nu$ and $\chi$ are added to the SM. The particle and scalar contents of the model and their assignments under the considered symmetries are summarized in Table \ref{particlecontS4} where we
define $\psi_L=\left(\psi_{1L}, \psi_{2L}, \psi_{3L}\right)^T$, $l_{R}=\left(l_{2R}, l_{3R}\right)^T$, $\nu_R=\left(\nu_{1R}, \nu_{2R}\right)^T$, $S=\left(S_1, S_2\right)^T$, $\varphi_{l}=\left(\varphi_{1l}, \varphi_{2l}, \varphi_{3l}\right)^T,  \phi_{l}=\left(\phi_{1l}, \phi_{2l}, \phi_{3l}\right)^T$ and $\varphi_{\nu}=\left(\varphi_{1\nu}, \varphi_{2\nu}, \varphi_{3\nu}\right)^T$ as multiplets of $S_4$.
\begin{table}[ht]
\caption{\label{particlecontS4}Particle content and their charge assignments under $SU(2)_L\times U(1)_Y\times S_4 \times Z_5\times Z_3\times Z_2$ $\, \big(\rho=e^{i\frac{2\pi}{5}},\, \om=e^{i\frac{2\pi}{3}}\big)$.}
	\centering
\begin{tabular}{@{}llllllllllllllllllll@{}}
\toprule
	    & $\psi_L$ & $l_{1R}$ &$l_{R}$ & $\nu_R$ & $S$ &$H$& $\varphi_l$ & $\phi_l$ & $\varphi_\nu$ &$\chi$ \\
	\midrule 
$\big[SU(2)_L, U(1)_Y\big]$& $\big[2, -\frac{1}{2}\big]$ &$\big[1, -1\big]$ &$\big[1, -1\big]$&$\big[1, 0\big]$&$\big[1, 0\big]$&$\big[2, \frac{1}{2}\big]$& $\big[1, 0\big]$ & $\big[1, 0\big]$&$\big[1, 0\big]$ &$\big[1, 0\big]$  \\
$S_4$ & 3$_1$ & 1$_1$ & 2& $2$ & $2$& $1_1$& 3$_1$ & 3$_2$& 3$_1$& 1$_1$ \\ 
$Z_5$ & $\rho^3$ & $\rho^2$& $\rho^2$& $\rho^2$ & $\rho$ &$\rho$& $1$ & $1$& $\rho^{2}$ & $\rho$ \\
$Z_3$ & 1 & $\omega$& $\omega$  & $\om^2$ & $\om$ & $\om^2$ & $1$ & $1$ & $1$& $\om$  \\
$Z_2$ & $-$ & $+$ & $+$ & $+$ & $-$ & $-$  & $+$ & $+$ & $+$  & $-$ \\  \toprule
	\end{tabular}
\end{table}\\
The given particle content yields the following 5D Yukawa terms:
\bea
-\mathcal{L} &=&\frac{h_1}{\Lambda}\left(\overline{\psi}_L \varphi_l\right)_{\mathbf{1}_1} \left(H l_{1R}\right)_{\mathbf{1}_1}
+ \frac{h_2}{\Lambda} \left(\overline{\psi}_L  \varphi_l\right)_{\mathbf{2}} \left(H l_R\right)_{\mathbf{2}}
+\frac{h_3}{\Lambda} \left(\overline{\psi}_L  \phi_l\right)_{\mathbf{2}} \left(H l_R\right)_{\mathbf{2}}  \crn
&+&\frac{x}{\Lambda} \left(\overline{\psi}_L \nu_R\right)_{\mathbf{3}_1}\big( \widetilde{H} \varphi_\nu\big)_{\mathbf{3}_1}
 + y\left(\overline{S} \nu_{R}\right)_{1_1} \chi^\ast
 +\frac{z}{2\Lambda}\left(\bar{S} S^c\right)_{\mathbf{1}_1} \chi^{2}  + h.c. \label{LlepS4}\eea
Here $\widetilde{H} = i \tau_2 H$, $\Lambda$ being the cut-off scale, $h_i\, (i=1\div 3), x, y$ and $z$ are the Yukawa-like couplings. Each of symmetries $Z_5, Z_3$ and $Z_2$ serves a crucial role in preventing the unwanted mass terms,
listed Appendix \ref{forbidappen}, to get the desired mass matrices.

The VEVs of scalar fields determined by the scalar potential minimum condition (see Appendix \ref{Higgspotential}) get the following forms:
\bea
&&\langle H\rangle =\left(0\hs\hs v\right)^T, \hs \langle \varphi_l \rangle = (v_{\varphi}, 0, 0), \hs \langle \phi_l \rangle = (v_{\phi}, 0, 0),\hs
\langle \varphi_\nu \rangle =(v_{1}, \,\, v_{2}, \,\, v_{3}), 
\hs \langle\chi\rangle =v_\chi.\label{VEV}
\eea

\section{\label{neutrinomixing} Neutrino mass and mixing}
Using the tensor product rules of $S_4$ in the $T$-diagonal basis \cite{HagedornS4basic,Kobayashi22}, from the first line of Eq. (\ref{LlepS4}), when the scalar fields $\phi_l, \varphi_l$ and $H$ obtain their VEVs in Eq. (\ref{VEV}), we get the following charged-lepton mass matrix\Red{,}
\bea
&&M_l=\mathrm{diag}\left(\frac{v}{\Lambda} h_1 v_{\varphi},\, \frac{v}{\Lambda}\big(h_2 v_{\varphi}-h_3 v_{\phi}\big),\, \frac{v}{\Lambda}\big(h_2 v_{\varphi} +h_3 v_{\phi}\big)\right) \equiv \mathrm{diag}\left(m_e,\, m_\mu,\, m_\tau\right). \label{Ml}\eea
The corresponding diagonalization matrices are therefore identity ones, $V_{eL}=V_{eR}=\mathbf{I}_{3\times 3}$, i.e., the charged leptons by themselves are the physical mass eigenstates and the lepton mixing matrix is fully that of neutrino.

Next, we consider the neutrino sector. With the aid of $S_4$ tensor products 
\cite{HagedornS4basic,Kobayashi22}, from the second line of Eq. (\ref{LlepS4}), after symmetry breaking, the mass Lagrangian for the neutrinos can be rewritten in the form
\bea
-\mathcal{L}^{mass}_\nu&=&\bar{\nu}_L M_{D}\nu_{R}+ \bar{S} M_{R} \nu_{R}+ \frac{1}{2}\mu \bar{S} S^C+h.c.
\equiv\frac{1}{2} \overline{n^C_L} M_\nu n_L +h.c, \eea
where
\bea
&&n_L =\left(\nu^C_{L} \hs\hs \nu_{R} \hs\hs S^C \right)^T, \hs
M_\nu =\left(%
\begin{array}{ccc}
0 &\,\,\, M_{D} &\,\,\,  0 \\
 M^T_{D} &  0 &\,\,\, M^T_{R} \\
0 & \,\,\, M_{R}  & \mu\\
\end{array}%
\right), \label{Mnu}\\
&&M_D=\left(%
\begin{array}{ccc}
b_D&c_D\\
a_D&b_D\\
c_D&a_D\\
\end{array}%
\right), \hs M_R=\left(%
\begin{array}{ccc}
0 & a_R\\
a_R & 0
\end{array}%
\right), \hs \mu=\left(%
\begin{array}{ccc}
0&a_\mu\\
a_\mu&0\\
\end{array}%
\right), \label{MDRmu}
\eea
with
\bea
a_D =
x v \left(\frac{v_{1} }{\Lambda}\right),\hs b_D = 
x v \left(\frac{v_{2} }{\Lambda}\right), \hs  c_D =
x v \left(\frac{v_{3} }{\Lambda}\right), \hs  a_R = y v_\chi,\hs  a_\mu = 
z v_{\chi} \left(\frac{v_{\chi} }{\Lambda}\right).
\label{aDRM}\eea
It is important to noted that although the matrix $M_D$ in our work is the same as that of Ref. \cite{ThapaJHEP23}, the matrices $M_R$ and $\mu$ are complete different from each other. Namely, in our model, $M_R$ and $\mu$ with zero diagonal elements and non-zero off-diagonal elements are naturally obtained 
whereas the corresponding matrix in Ref. \cite{ThapaJHEP23} is obtained by assuming the Yukawa coupling constants in the interaction between two SM gauge singlet fermions ($S_{1}$ and $S_2$) and two right-handed neutrinos ($N_R$) are the same, $\gamma_1=\gamma_2$. On the other hand, the matrix $\mu$ in Ref. \cite{ThapaJHEP23} has diagonal form and 
obtained by assuming the Yukawa coupling constants in the Majorana mass terms of sterile neutrinos are the same, $\lambda_1=\lambda_2$ whereas in our model $\mu$ is naturally obtained due to the symmetry of $1_1$ as a result of $2\times 2$ 
of $S_4$. Besides, in our model, $\mu/M_R\propto v_{\chi}/\Lambda\ll 1$, i.e., the condition $\mu\ll M_R$ for the Inverse Seesaw Mechanism is naturally satisfied.

The comments are in order:
\begin{itemize}
  \item [$(i)$] Suposing that the Yukawa couplings in neutrino sector\footnote{The electroweak symmetry breaking scale is
      $v=246\, \mathrm{GeV}$.} are $x\sim z \sim \mathcal{O}(1), \, y\sim \mathcal{O}(10^{-1})$, $v_{1}\sim v_{2}\sim v_{3} \sim 10^{11}\, \mathrm{GeV}$ and $v_{\chi} \sim 10^{5}\, \mathrm{GeV}$; thus, with $\Lambda\sim 10^{13}$ GeV we can estimate $\mu\sim 10^{-3}$ GeV, $M_D\sim 1$\, GeV, and $M_R\sim 10^4$ GeV, i.e., $\mu\ll M_D\ll M_R$. Therefore, the mass of light neutrinos can be obtained via the ISS mechanism, 
      \bea
      m_\nu = M_D \left(M_R^T\right)^{-1} \mu \left(M_R\right)^{-1} M_D^T. \label{mnuISS}
      \eea
      With the aid of Eq. (\ref{mnuISS}),
      $m_\nu \sim 10^{-2}$ eV may be achieved by the scales of $\mu, M_D$ and $M_R$.
  \item [$(ii)$] The mass scale of the heavy
neutrinos $M_R\sim 10^4$ GeV in the considered model is much lower than that of the canonical seesaw making it can be tested by future colliders \cite{Mekałaplb23,Lijhep23,Hongprd2024,Shenaepjc25}.
\end{itemize}
Substituting  Eq. (\ref{MDRmu}) into Eq. (\ref{mnuISS}) yields
\bea
	m_\nu =m\left(
\begin{array}{ccc}
 2 \beta & \al+\beta^2 & \al \beta+1 \\
 \al+\beta^2 & 2 \al \beta & \al^2+\beta \\
 \al \beta+1 & \al^2+\beta & 2 \al \\
\end{array}
\right), \label{Mnu}
\eea
where
\bea
&&m=\frac{c_D^2 a_\mu}{a_R^2}, \hs \alpha = \frac{a_D}{c_D}, \hs \beta = \frac{b_D}{c_D}. \label{m0ab}
\eea
It is noted that $m, \alpha$ and $\beta$ are three complex parameters 
where $m$ has the dimension of mass while $\alpha$ and $\beta$ are dimensionless.

The Yukawa couplings $x, y$  and $z$ are, in general, complex parameters, thus, $a_D, b_D, c_D, a_R$ and $a_\mu$ are complex, then $m_\nu$ in Eq. (\ref{Mnu}) is a complex matrix. The light neutrinos masses are obtained by diagonalising the Hermitian matrix,
\begin{equation}
h=m_{\nu}m_{\nu}^{\dagger}=m_0^2\begin{pmatrix}
\mathbf{a} & \mathbf{b} & \mathbf{c} \\
\mathbf{b}^* & \mathbf{d} & \mathbf{g} \\
\mathbf{c}^*& \mathbf{g}^* & \mathbf{f}
\end{pmatrix} \hspace{0.35 cm} (m_0=|m|),
\label{hmatrix}
\end{equation}
where
\begin{align}
&\mathbf{a} =\vert\alpha+\beta^2\vert^2 +\vert\alpha \beta+1\vert^2 + 4 \vert \beta \vert^2, \nonumber\\
&\mathbf{b} = 2 \left(\alpha+\beta^2\right) (\alpha \beta)^*+(\alpha \beta+1) \big[\left(\alpha^*\right)^2+\beta^*\big]+2 \beta \big[\alpha^*+\left(\beta^*\right)^2\big], \nonumber\\
&\mathbf{c} = \left(\alpha+\beta^2\right) \big[\left(\alpha^*\right)^2+\beta^*\big]+2 (\alpha \beta+1) \alpha^*+2 \beta \big[(\alpha \beta)^*+1\big], \nonumber\\
&\mathbf{d} =\vert\alpha^2+\beta\vert^2 + \vert\alpha+\beta^2\vert^2 + 4 \vert\alpha \beta\vert^2,   \nonumber\\
&\mathbf{g} = 2 \left(\alpha^2+\beta\right) \alpha^*+\left(\alpha+\beta^2\right) \big[(\alpha \beta)^*+1\big]+2 \alpha \beta \big[\left(\alpha^*\right)^2+\beta^*\big],  \nonumber\\
&\mathbf{f} = \vert\alpha^2+\beta\vert^2 +\vert\alpha \beta+1\vert^2 + 4 \vert\alpha\vert^2.\label{abcdgf}
\end{align}
The matrix $h$ in Eq. (\ref{hmatrix}) can be diagonalized by the PMNS mixing matrix $U_{\textrm{PMNS}}$,
\bea
U_{\textrm{PMNS}}^\dagger h U_{\textrm{PMNS}}^* =\left\{
\begin{array}{l}
m^2_1 = 0, \,\, m_{2,3}^2 = m_0^2\left(\kappa_0 \mp 2 \sqrt{\kappa_1}+\kappa_2\right)
\hspace{0.15 cm} \mbox{for NH,}\ \  \\
m^2_3 = 0, \,\, m_{1,2}^2 = m_0^2\left(\kappa_0 \mp 2 \sqrt{\kappa_1}+\kappa_2\right)
\hspace{0.15 cm} \mbox{for IH,}
\end{array}
\right. \label{bestfitpoints}
\eea
where \begin{align}
\kappa_0 = &\ 1 + \alpha  \beta +\vert \alpha \vert ^4+\vert \beta \vert ^4, \nonumber\\
\kappa_1=&\left(\alpha +\beta  \alpha ^*+\beta ^*\right) \left(\beta +\alpha  \beta ^*+\alpha ^*\right) \left(\vert \alpha \vert^2+\vert\beta \vert ^2+1\right)^2 , \nonumber\\
\kappa_2 =&\left(\alpha ^2+3 \beta +\alpha  \beta ^*\right) \beta ^*  + \left(3 \alpha +\beta ^2+3 \alpha  \beta  \beta ^*+\beta ^*\right) \alpha ^* + \left(\alpha ^*\right)^2 \beta. \label{kapa012}
\end{align}
In standard parametrization, $U_{\textrm{PMNS}}$ is given by
\bea
	U_{\textrm{PMNS}} =
	\begin{pmatrix}
	c_{12} c_{13} & s_{12} c_{13} & s_{13} e^{-i \delta_{CP}} \\
	-s_{12}c_{23}-c_{12}s_{23}s_{13}e^{i\delta_{CP}} & c_{12}c_{23}-s_{12}s_{23}s_{13}e^{i\delta_{CP}} & s_{23}c_{13} \\
	s_{12}c_{23}-c_{12}c_{23}s_{13}e^{i\delta_{CP}} & -c_{12}s_{23}-s_{12}c_{23}s_{13}e^{i\delta_{CP}} & c_{23}c_{13}
	\end{pmatrix}
	P, \label{UPMNSpara}
\eea
where, in this work, the lightest neutrino mass $m_{light}=m_1=0$ for NH and $m_{light}=m_3=0$ for IH; thus, $P=\mathrm{diag}(1, e^{i\sigma}, 1)$.

Three neutrino mixing angles $\theta_{12},\theta_{13}, \theta_{23}$ and Dirac CP violating phase $\delta_{CP}$ are expressed in terms of the model parameters as \cite{Xingparameters,ThapaJHEP23}\Blue{,}
\begin{align}
&\tan\theta_{23}=  \frac{\mathrm{Im} \mathbf{b}}{\mathrm{Im} \mathbf{c}}, \hs
\tan2\theta_{12}=  \frac{2 N_{12}}{N_{22}-N_{11}}, \nonumber\\
&\tan\theta_{13}=  \vert \mathrm{Im} \mathbf{g}\vert\cdot \frac{\sqrt{\left[(\mathrm{Im} \mathbf{b})^2 + (\mathrm{Im} \mathbf{c})^2\right]^2 + \left(\mathrm{Re} \mathbf{b}\mathrm{Im} \mathbf{b}+ \mathrm{Im} \mathbf{c} \mathrm{Im} \mathbf{c}\right)^2}}{\sqrt{\left[(\mathrm{Im} \mathbf{b})^2 + (\mathrm{Im} \mathbf{c})^2\right] \left(\mathrm{Re} \mathbf{b}\mathrm{Im} \mathbf{c}-\mathrm{Im} \mathbf{b} \mathrm{Re} \mathbf{c}\right)^2} }, \nonumber\\
&\tan\delta_{CP} =  - \frac{(\mathrm{Im} \mathbf{b})^2 + (\mathrm{Im} \mathbf{c})^2}{\mathrm{Re} \mathbf{b}\mathrm{Im}\mathbf{b}+ \mathrm{Re} \mathbf{c} \mathrm{Im} \mathbf{c}}, \label{Neutrino_parameter}
\end{align}

where the quantities $N_{11}, N_{12}$ and $N_{22}$ are defined by
\begin{align}
N_{11}=&\ \mathbf{a}- \frac{\mathrm{Re} \mathbf{b} \mathrm{Im} \mathbf{c}-\mathrm{Im} \mathbf{b} \mathrm{Re} \mathbf{c}}{\mathrm{Im} \mathbf{g}},   \nonumber \\
N_{12}=&\ \left[\frac{\big(\mathrm{Re} \mathbf{b} \mathrm{Im} \mathbf{c} - \mathrm{Im} \mathbf{b} \mathrm{Re} \mathbf{c}\big)^2}{(\mathrm{Im}\mathbf{b})^2 + (\mathrm{Im} \mathbf{c})^2} + \left(\frac{\big[\mathrm{Re} \mathbf{b} \mathrm{Im} \mathbf{b}+ \mathrm{Re}\mathbf{c}\mathrm{Im}\mathbf{c}\big]^2}{\big[(\mathrm{Im}\mathbf{b})^2 + (\mathrm{Im} \mathbf{c})^2\big]^2}+1 \right) (\mathrm{Im}\mathbf{g})^2    \right]^{\frac{1}{2}},  \label{Nij} \\
N_{22} =&\ \frac{(\mathrm{Im}\mathbf{c})^2 \mathbf{d} + (\mathrm{Im}\mathbf{b})^2 \mathbf{f} - 2 \mathrm{Im} \mathbf{b} \mathrm{Im} \mathbf{c}\mathrm{Re}\mathbf{g}}{(\mathrm{Im}\mathbf{b})^2 + (\mathrm{Im} \mathbf{c})^2}. \nonumber
\end{align}
With the aid of Eqs. (\ref{bestfitpoints}) and (\ref{kapa012}), the sum of neutrino masses $\sum=\sum_{i=1}^{3} m_i$ can be expressed in terms of $m_0$, $\alpha$ and $\beta$ as follows
\bea
\sum&=&m_0\Big(\sqrt{\kappa_0 + 2 \sqrt{\kappa_1}+\kappa_2} + \sqrt{\kappa_0 - 2 \sqrt{\kappa_1}+\kappa_2}\Big).\label{sumnuexpres}
\eea
The effective Majorana neutrino mass, $m_{\beta\beta} =\left|\sum^3_{k=1} \big(U_{1 k}\big)^2 m_k \right|$, is given by
\bea
&&m_{\beta\beta} = m_0\left|c_{13}^2 s_{12}^2 \sqrt{\kappa_0 - 2 \sqrt{\kappa_1}+\kappa_2}\, e^{2i \sigma}+s_{13}^2 \sqrt{\kappa_0 + 2 \sqrt{\kappa_1}+\kappa_2}\, e^{-2i\delta_{CP}}\right|, \label{mbb}\eea
where $\kappa_0, \kappa_1$ and $\kappa_2$ are given in Eq. (\ref{kapa012}).
\section{\label{reslepto}Resonant leptogenesis}
In this section, we study the leptogenesis in the ISS(2,2) framework. In contrast to conventional thermal leptogenesis, where the CP-violating parameter is suppressed by large mass differences between heavy Majorana states,
the inverse seesaw mechanism naturally features pairs of quasi-degenerate heavy neutrinos, whose small mass splitting induced by the lepton-number-violating parameter $\mu$, can lead to a resonant enhancement of the CP asymmetry.
The $4\times 4$ block of heavy neutrino mass matrix in the basis ($\nu_R , S$) takes the
form
\begin{align}
M_{\nu S} = \left(\begin{array}{cc}
0 & M_R \\
M_R^T & \mu
\end{array} \right), \label{eq:Mnus}
\end{align}
where $M_R$ and $\mu$ are $2\times 2$ real symmetric matrices, defined in Eq. (\ref{MDRmu}).

Diagonalising the mass matrix Eq.~(\ref{eq:Mnus}) via a unitary matrix $U_{\mathrm{mass}}$,
\begin{equation}
U_{\mathrm{mass}}^T M_{\nu S} U_{\mathrm{mass}} = \mathrm{diag}(M_1, M_2, M_3, M_4).
\end{equation}

yields four mass eigenvalues
\begin{align}
M_1 &= \tfrac12\!\left(-a_\mu - \sqrt{4a_R^2 + a_\mu^2}\right), \hspace{0.35cm}
M_2 = \tfrac12\!\left( a_\mu - \sqrt{4a_R^2 + a_\mu^2}\right), \nonumber\\
M_3 &= \tfrac12\!\left(-a_\mu + \sqrt{4a_R^2 + a_\mu^2}\right), \hspace{0.35cm}
M_4 = \tfrac12\!\left( a_\mu + \sqrt{4a_R^2 + a_\mu^2}\right).
\end{align}
Physical masses are defined as $M_i=|M_i|$, since negative eigenvalues can be absorbed by field redefinitions.

In the phenomenologically relevant limit $a_\mu \ll a_R$, the mass spectrum consists of two quasi-Dirac pairs,
\begin{equation}
M_\pm \simeq a_R \pm \frac{a_\mu}{2},
\end{equation}
where each pair splits by a small lepton number violating mass difference controlled by the parameter $a_\mu$. Resonant enhancement of the CP asymmetry occurs within each quasi-Dirac pair when the mass splitting becomes comparable to the decay widths of the heavy states.

The Dirac neutrino mass matrix in the ISS(2,2) model is constructed using the Casas-Ibarra parametrization,
\begin{equation}
m_D = U_{\mbox{PMNS}} m_d^{1/2} R \mu^{-1/2} M_R^T, \label{mDCI}
\end{equation}
where, $m_d = \mathrm{diag}(m_1, m_2, m_3)$ is the diagonal active neutrino mass
matrix and $R$ is a complex $3\times 2$ orthogonal matrix satisfying $R^T R = \mathbb{I}$, given by

\begin{equation}
R = \left(\begin{matrix}
0 & 0 \\
\cos \zeta & \sin\zeta \\
-\sin \zeta & \cos \zeta
\end{matrix}\right),
\end{equation}
where $ \zeta= \mathrm{Re}[\zeta] + i \mathrm{Im}[\zeta]$.

It is noted that the one loop correction to neutrino mass is given as \cite{Grimusplb2002}, $M_{\nu L}^{\mathrm{1-loop}} \simeq \frac{f(x_N)}{m_W^2}m_D \mu m_D^T$, where the one loop function $f(x_N)$ is given by
$f(x_N)=
\frac{\alpha_W}{16\pi}
\Big[
\frac{m_H^2}{M_N^2 - m_H^2}
\ln\!\Big(\frac{M_N^2}{m_H^2}\Big)
+
3\,\frac{m_Z^2}{M_N^2 - m_Z^2}
\ln\!\Big(\frac{M_N^2}{m_Z^2}\Big)
\Big]$, with $a_W = g^2/4\pi$ and $ M_N = \frac{1}{N}\sum M_i$ is the average heavy neutrino mass, and $m_H, m_Z$ are the Higgs and $Z$-boson masses. 
In the limit $M_N^2 \gg m_W^2, $ the loop function is reduced to
$f(x_N)
\simeq
\frac{\alpha_W}{16\pi}
\Big(
\frac{m_H^2 + 3 m_Z^2}{M_N^2}
\Big)
\ln\!\Big(\frac{M_N^2}{m_W^2}\Big)$ and
$M_{\nu L}^{\mathrm{1-loop}}
\simeq
\Delta_{\text{rad}} \,
m_\nu^{\text{tree}}$.
In our analysis, $M_N = 1-10$ TeV, we obtain
$\Delta_{\text{rad}} \sim 0.02 - 0.04$.
Therefore, even for large values of $\mathrm{Im}[\zeta]$, fine-tuned cancellation is not involved and the light neutrino masses remain stable under radiative corrections.

In the interaction basis, only the two right-handed neutrinos couple to the SM lepton doublets. The full Yukawa matrix therefore has the form
\begin{equation}
h_I = \frac{\sqrt{2}}{v}\left(m_D
\;\;\;
0_{3\times2}
\right),
\end{equation}
which is a $3\times4$ matrix. The Yukawa couplings in the mass basis are given as
\begin{equation}
h_{\alpha i} = (h_I\hs  U_{\mathrm{mass}})_{\alpha i}.
\end{equation}

The flavoured CP asymmetry parameter for the decay of a heavy neutrino $N_i$ into a lepton of flavour $\alpha$ is defined as \cite{Pilaftsisnpb04}
\begin{equation}
\varepsilon_{\alpha i} \;=\;
\frac{\Gamma(N_i \to \ell_\alpha H) - \Gamma(N_i \to \bar{\ell}_\alpha H^\dagger)}
     {\sum_\beta \big[ \Gamma(N_i \to \ell_\beta H) + \Gamma(N_i \to \bar{\ell}_\beta H^\dagger) \big]} \, . \label{varepial}
\end{equation}
In terms of Yukawa couplings $h_{\alpha i}$, 
expression (\ref{varepial}) can be written as \cite{Pilaftsisnpb04, Chakraborty, Iso,Garny}
\begin{equation}
\varepsilon_{\alpha i}
=
\frac{1}{8 \pi (h^\dagger h)_{ii}}
\sum_{j\neq i =1}^{4}
\mathrm{Im}
\!\left[
h_{\alpha i}^\ast h_{\alpha j}(h^\dagger h)_{ij}
\right]
\frac{(M_i^2-M_j^2)M_iM_j}
{(M_i^2-M_j^2)^2 + (M_i\Gamma_i+M_j\Gamma_j)^2}.
\end{equation} 
The total decay width of $N_i$ is given by
\begin{equation}
\Gamma_i = \frac{M_i}{8\pi}\,(h^\dagger h)_{ii}.
\end{equation}
Although the heavy spectrum is grouped into two quasi-Dirac pairs, the diagonalisation of the heavy neutrino mass matrix gives four distinct Majorana states with generally different Yukawa couplings in the mass basis. We therefore treat the decays of all four heavy states explicitly in the Boltzmann equations. Defining the heavy neutrino yields \(Y_{N_i}=n_{N_i}/s\) and the flavoured lepton asymmetries \(\Delta_\alpha\) (\(\alpha=e,\mu,\tau\)), the Boltzmann equations are defined as \cite{Chakraborty, Plumacher},
\begin{align}
&\frac{dY_{N_i}}{dz} = - \frac{z}{sH}
\gamma_{D_i}
\left(
\frac{Y_{N_i}}{Y_{N_i}^{\rm eq}} - 1
\right),
\qquad i=1,\dots,4, \\
&\frac{d\Delta_\alpha}{dz}=\frac{z}{sH}
\left[
\sum_{i=1}^{4}
\varepsilon_{\alpha i}\,
\gamma_{D_i}
\left(
\frac{Y_{N_i}}{Y_{N_i}^{\rm eq}} - 1
\right)
-
\left(
\sum_{i=1}^{4}
P_{\alpha i}\,\gamma_{D_i}
\right)
\frac{\Delta_\alpha}{Y_\ell^{\rm eq}}
\right],
\end{align}
where the equilibrium lepton yield is
\begin{equation}
Y_\ell^{\rm eq} = \frac{15}{4\pi^2 g_*}.
\end{equation}
The flavour projectors are defined as
\begin{equation}
P_{\alpha i}
=
\frac{|h_{\alpha i}|^2}{(h^\dagger h)_{ii}},
\qquad
\sum_\alpha P_{\alpha i}=1 .
\end{equation}
In numerical analysis, we solve the Boltzmann equations using a common evolution variable
\begin{equation}
z \equiv \frac{M_0}{T},
\qquad
M_0 \equiv \mathrm{min}(M_i) ,
\end{equation}
where \(M_i\) are the physical heavy-neutrino masses obtained after numerical  diagonalisation. The temperature at a given value of \(z\) is therefore
\begin{equation}
T(z) = \frac{M_0}{z}.
\end{equation}
For each heavy state we further define
\begin{equation}
z_i \equiv \frac{M_i}{T} = \frac{M_i}{M_0}\,z,
\qquad i=1,\dots,4 .
\end{equation} 
The entropy density and Hubble expansion rate are given by
\begin{equation}
s(T) = \frac{2\pi^2}{45}\,g_*\,T^3,
\qquad
H(T) = 1.66\,\sqrt{g_*}\,\frac{T^2}{M_{\rm Pl}} .
\end{equation} 
The equilibrium yield and number density of each heavy neutrino are
\begin{align}
&Y_{N_i}^{\rm eq}
=
\frac{45}{4\pi^4}\frac{g}{g_*}
z_i^2 K_2(z_i)
\qquad\hs (g=2), \\ 
&n_{N_i}^{\rm eq}
=
\frac{g M_i^2 T}{2\pi^2} K_2(z_i).
\end{align}
The thermally averaged decay reaction density is
\begin{equation}
\gamma_{D_i}
=
n_{N_i}^{\rm eq}
\frac{K_1(z_i)}{K_2(z_i)}\,\Gamma_i .
\end{equation}
The system is evolved numerically from \(z=10^{-2}\) to \(z=100\). At the final integration point, the total \(B-L\) asymmetry is obtained as
\begin{equation}
Y_{B-L} = \sum_{\alpha=e,\mu,\tau} \Delta_\alpha(z=100).
\end{equation}
Due to the sign convention adopted in the Boltzmann equations, the physical asymmetry is given by \(Y_{B-L}^{\rm phys} = -\,Y_{B-L}\).
Electroweak sphaleron processes convert this into baryon number according to
\begin{equation}
Y_B = \frac{28}{79}\,Y_{B-L}^{\rm phys}.
\end{equation}
Finally, the baryon-to-photon ratio is computed as
\begin{equation}
\eta_B=7.04\,Y_B.
\end{equation}

\section{\label{CLV}Charged Lepton Flavour Violation}

We now show how the ISS(22) mechanism can contribute to cLFV processes such as
$\mu \to \gamma e$, $\tau \to \gamma e$ and $\tau \to \gamma \mu$ at one-loop level due to the exchange of heavy
neutrinos and the $W$ boson. The branching ratio is given by~\cite{Hisano:1995cp,Ilakovac:1994kj,AbadaDas,AbadaCLFV,VallecLFV}

\begin{equation}
\text{BR}(\ell_i \to \ell_j \gamma) =
\frac{\alpha_{\text{em}}^3 \, \sin^2\theta_W}{256\pi^2 m_W^4} \,
\frac{m_{\ell_i}^5}{\Gamma_{\ell_i}}
\left| \sum_k U_{i k} U^{*}_{j k} \,
F\!\left(\frac{M_k^2}{m_W^2}\right) \right|^2 ,
\label{eq:cLFV-BR}
\end{equation}
\begin{equation}
\text{BR}(\ell_{\alpha} \to \ell_{\beta} \gamma) =
\frac{\alpha_{\text{em}}^3 \, \sin^2\theta_W}{256\pi^2 m_W^4} \,
\frac{m_{\ell_{\alpha}}^5}{\Gamma_{\ell_{\alpha}}}
\left| \sum_{i} U_{\alpha i} U^{*}_{\beta i} \,
F\!\left(\frac{M_{i}^2}{m_W^2}\right) \right|^2 ,
\label{eq:cLFV-BR}
\end{equation}
where $m_{\ell_{\alpha}}$ and $\Gamma_{\ell_{\alpha}}$ are the masses and decay widths of the initial leptons,
and $M_{i}$ are the masses of the heavy neutrinos. The branching ratios depend on the fine structure constant ($\alpha_{\text{em}}$), 
$W$ boson mass ($m_W$) and Weinberg angle $(\theta_W)$. The decay widths of the initial leptons are measured as $\Gamma_{\mu} = 2.996\times 10^{-19}$ for muons and $\Gamma_{\tau} = 2.267 \times 10^{-12}$ for tau leptons. The heavy--light mixing is defined as
\begin{equation}
U_{\alpha i} \;\simeq\; \frac{v}{\sqrt{2}} \frac{h_{\alpha i}}{M_{i}} \, ,
\end{equation}
with $v=246$~GeV and $h_{\alpha i}$ is the Yukawa couplings.
The loop function $F(x)$, with $x=\frac{M_{i}^2}{m_W^2}$,
has the form, 
\bea
F(x) &=& \frac{1}
{6 (1-x)^4} \big[4 x^4 + (18 \ln x - 49) x^3 + 78 x^2 - 43 x + 10\big].
\label{eq:loopF}
\eea

\begin{figure}
\begin{center}
\hspace*{2.0 cm}
\subfigure{\includegraphics[width=0.95\textwidth]{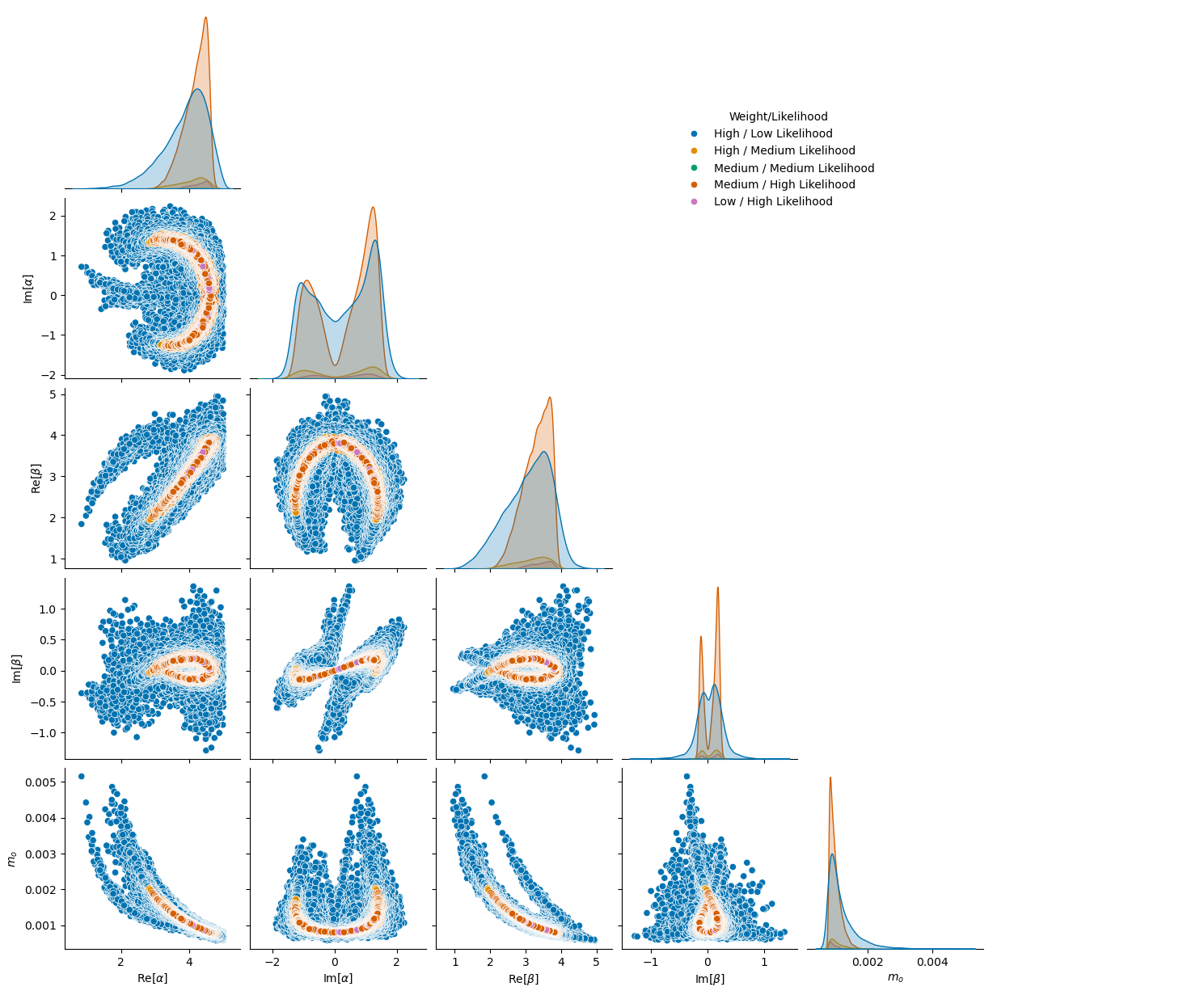}}\hspace{-0.45 cm}
\vspace{-0.25cm}
\caption{\footnotesize{Pairwise relationships between the neutrino model parameters. 
The values of likelihood and weight are shaded in different colours. These colours indicate the combination of weight (volume in parameter space) and likelihood (quality of fit). High-weight points represent broad allowed regions, while high likelihood points corresponds to better fits to the experimental data. The orange colour indicate the free parameter values which fit the neutrino oscillation data with high accuracy.  }}
\label{logplot}
\end{center}
\vspace{-0.5cm}
\end{figure}

\section{\label{numericalana}Numerical analysis}
The squared light neutrino mass matrix in Eqs.(\ref{hmatrix}) and (\ref{abcdgf}) contains five  free real parameters $\mathrm{Re}\alpha, \mathrm{Im}\alpha, \mathrm{Re}\beta, \mathrm{Im}\beta$ and $m_0$. In order to fit the observed neutrino
data \cite{Salas2021} (see Table \ref{Salas2021T}), we use a $\chi^2$ function and perform a numerical simulation utilizing a Multinest sampling package \cite{Multinest}. Minimizing $\chi^2$ yields the best-fit values of the model parameters and the prediction of neutrino observables.
The $\chi^2$ is defined as
\begin{equation}
\chi^2(x_i) = \sum_{j}\left(\frac{y_j(x_i)-y_j^{bf}}{\sigma_j}\right)^2,
\label{chitest}
\end{equation}
where  $j$ is summed across the neutrino observables $\sin^2\theta_{12},\ \sin^2\theta_{13},\ \sin^2\theta_{23}, \Delta m_{21}^2$ and $\Delta m_{31}^2$, while $x_i$ are free parameters in the model; $y_j(x_i)$ are the model predictions for the observables 
while $y_j^{bf}$ and
$\sigma_j$ are respectively the best-fit points and the corresponding 3$\sigma$ uncertainties taken from Ref. \cite{Salas2021} (Table \ref{Salas2021T}).

We also define the parameter $r = \sqrt{\frac{\Delta m_{21}^2}{\Delta m_{31}^2}}= \frac{m_2}{m_3}$.
Furthermore, since the Dirac CP phase $\delta_{CP}$ has not been significantly restricted, it is not treated as an input parameter.
The free parameters of the model are randomly scanned in the following ranges
\begin{equation}
\mathrm{Re}\alpha,\, \mathrm{Im}\alpha,\,\mathrm{Re}\beta,\,\mathrm{Im}\beta \in [-5,5], \hspace{0.25 cm} m_{0} \in [0,1] \, \mathrm{meV.}
\label{freepara}
\end{equation}

Figure \ref{logplot} shows the pairwise relationships between the free model parameters based on the log-likelihood and weight for NH. We find a highly localized values of the parameters with smooth distribution curve in a narrow range. These plots show strong relationships between parameters, especially between Re$\beta$ and Im$\beta$.
\begin{figure}
\subfigure[]{
\includegraphics[width=.45\textwidth]{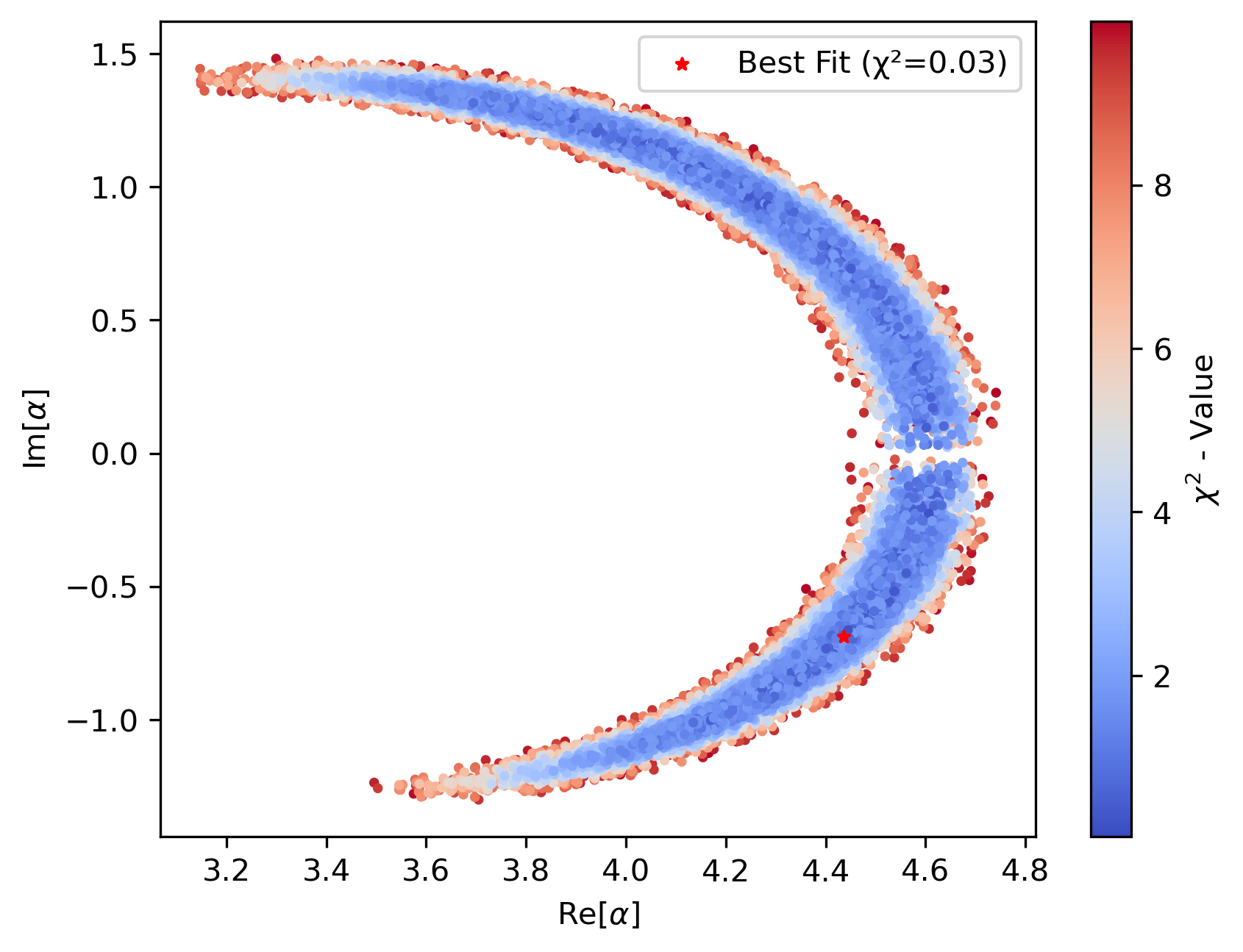}}
\quad
\subfigure[]{
\includegraphics[width=.45\textwidth]{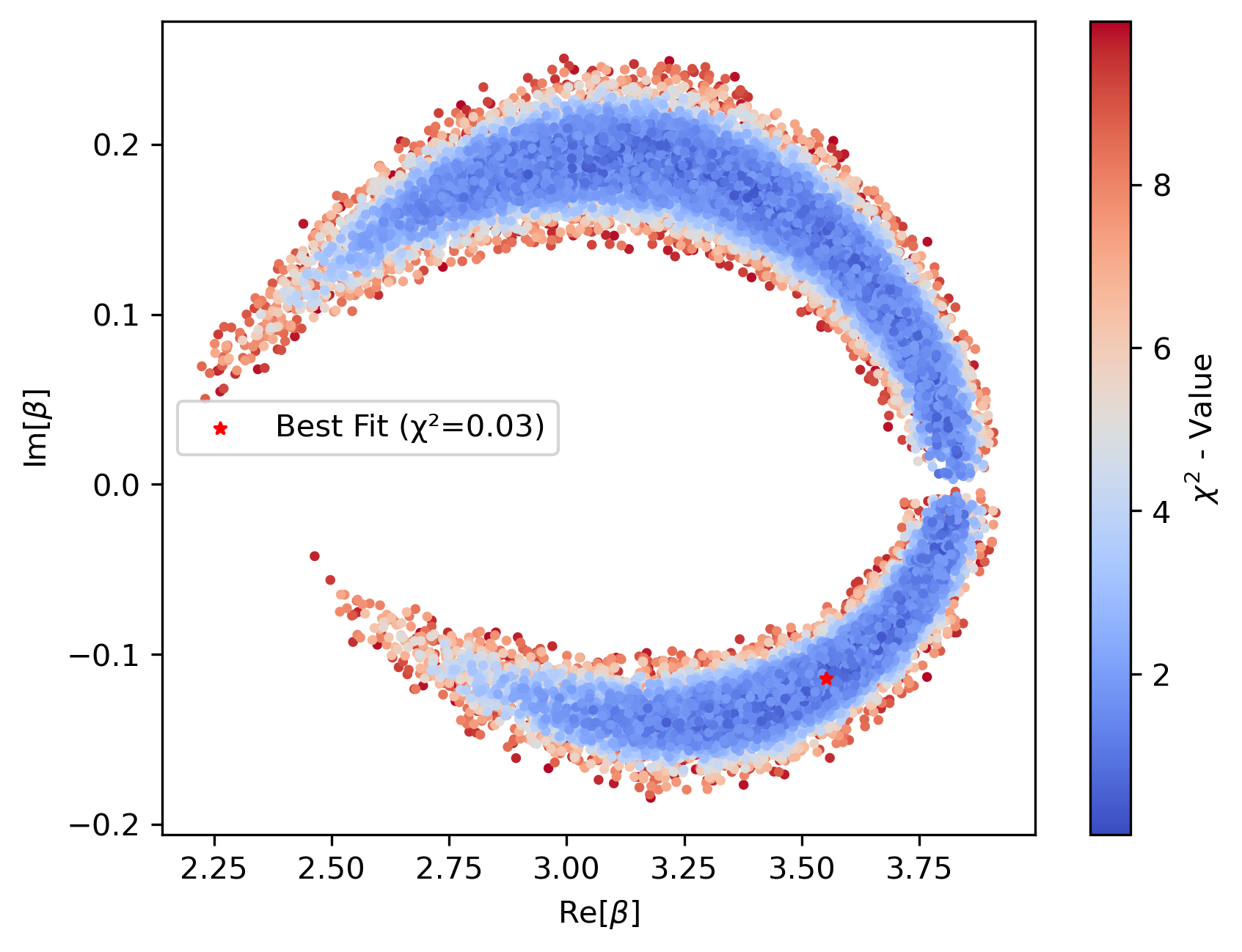}}
\quad
\vspace{-0.15 cm}
\caption{\footnotesize{Allowed regions of (a) Re[$\alpha$] and Im$[\alpha]$, and (b) of Re[$\beta$] and Im$[\beta]$.}}
\label{abplot}
\end{figure}
\begin{figure}
\subfigure[]{\includegraphics[width=.45\textwidth]{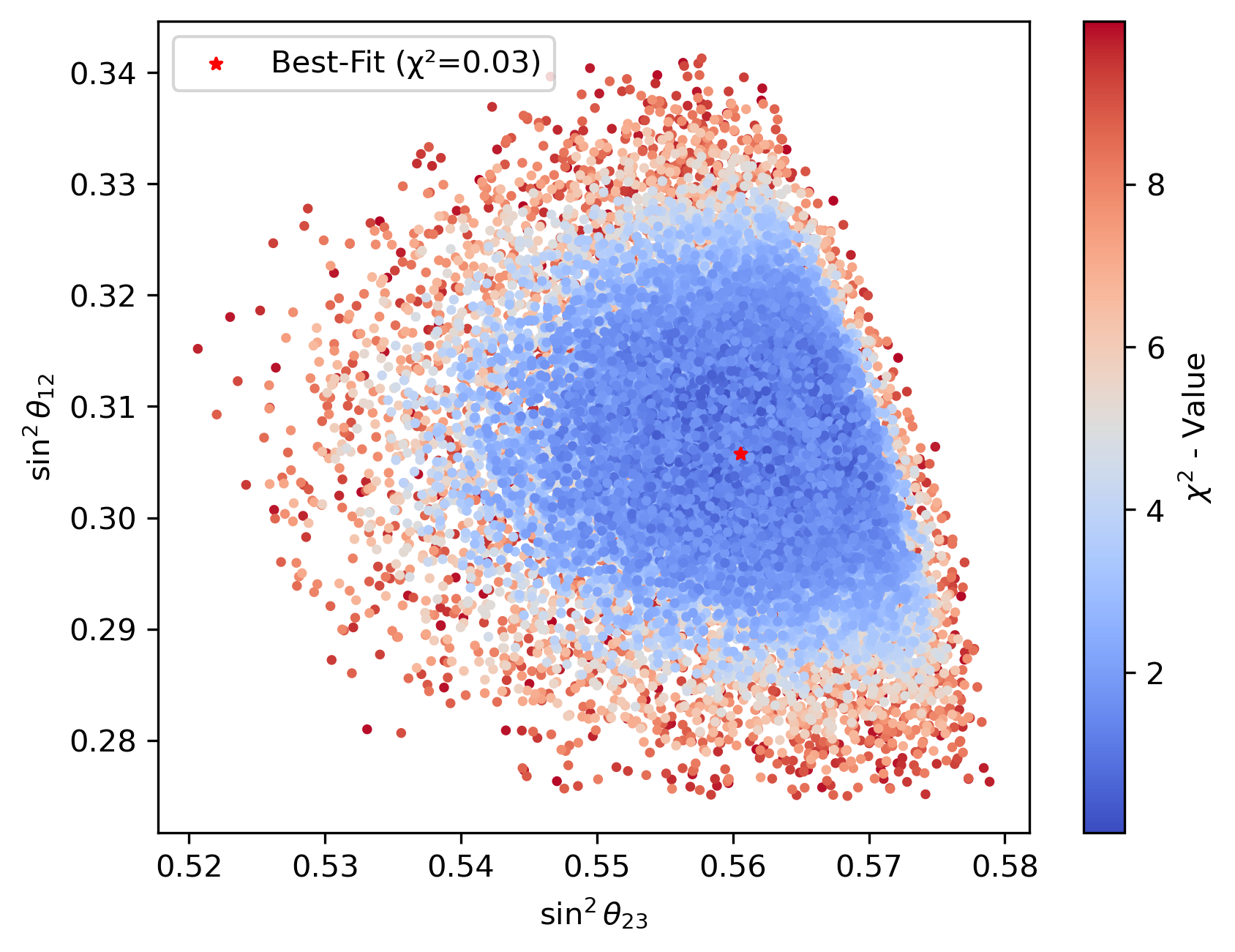}}
\quad
\subfigure[]{\includegraphics[width=.45\textwidth]{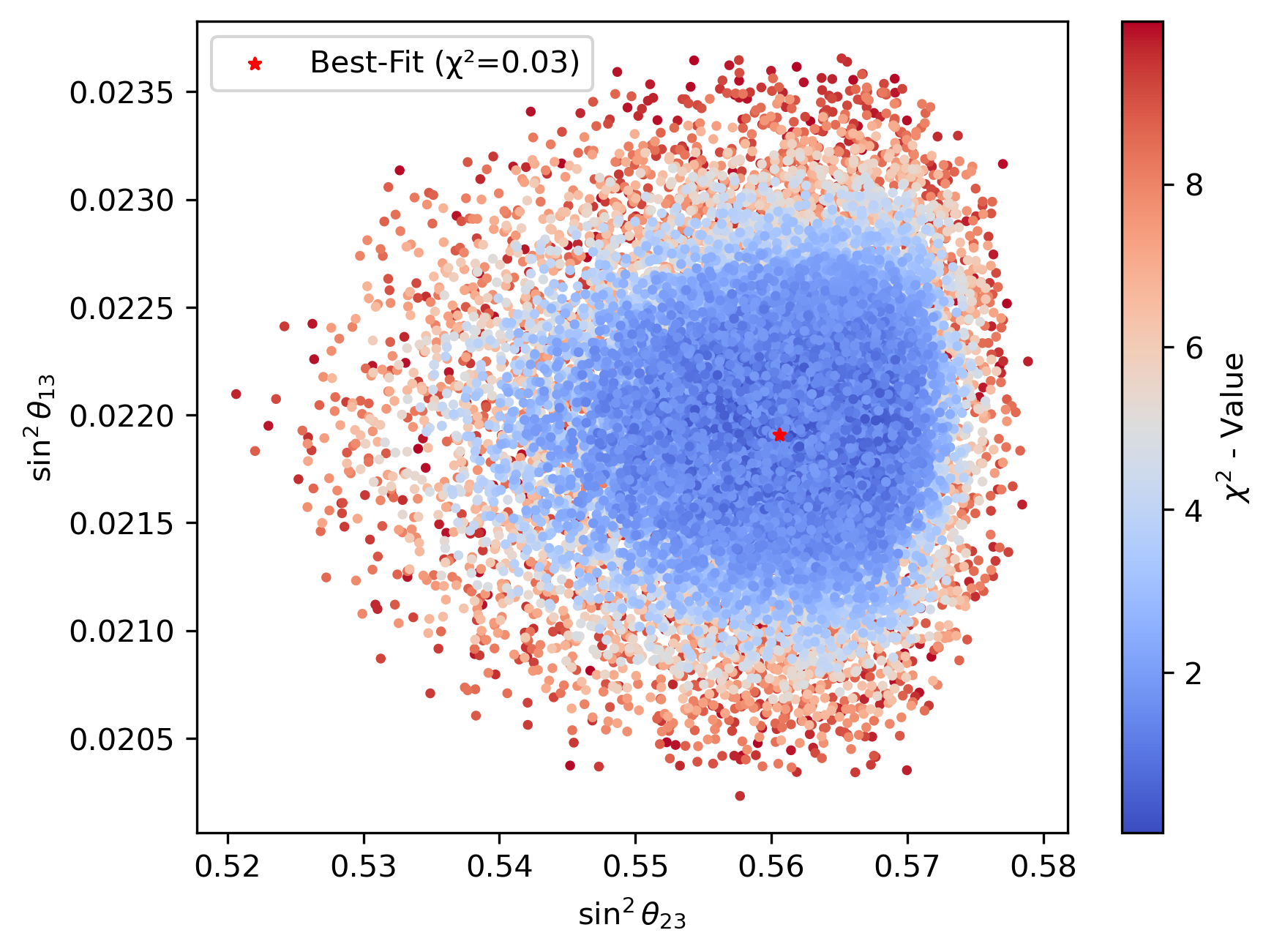}}
\quad
\subfigure[]{
\includegraphics[width=.47\textwidth]{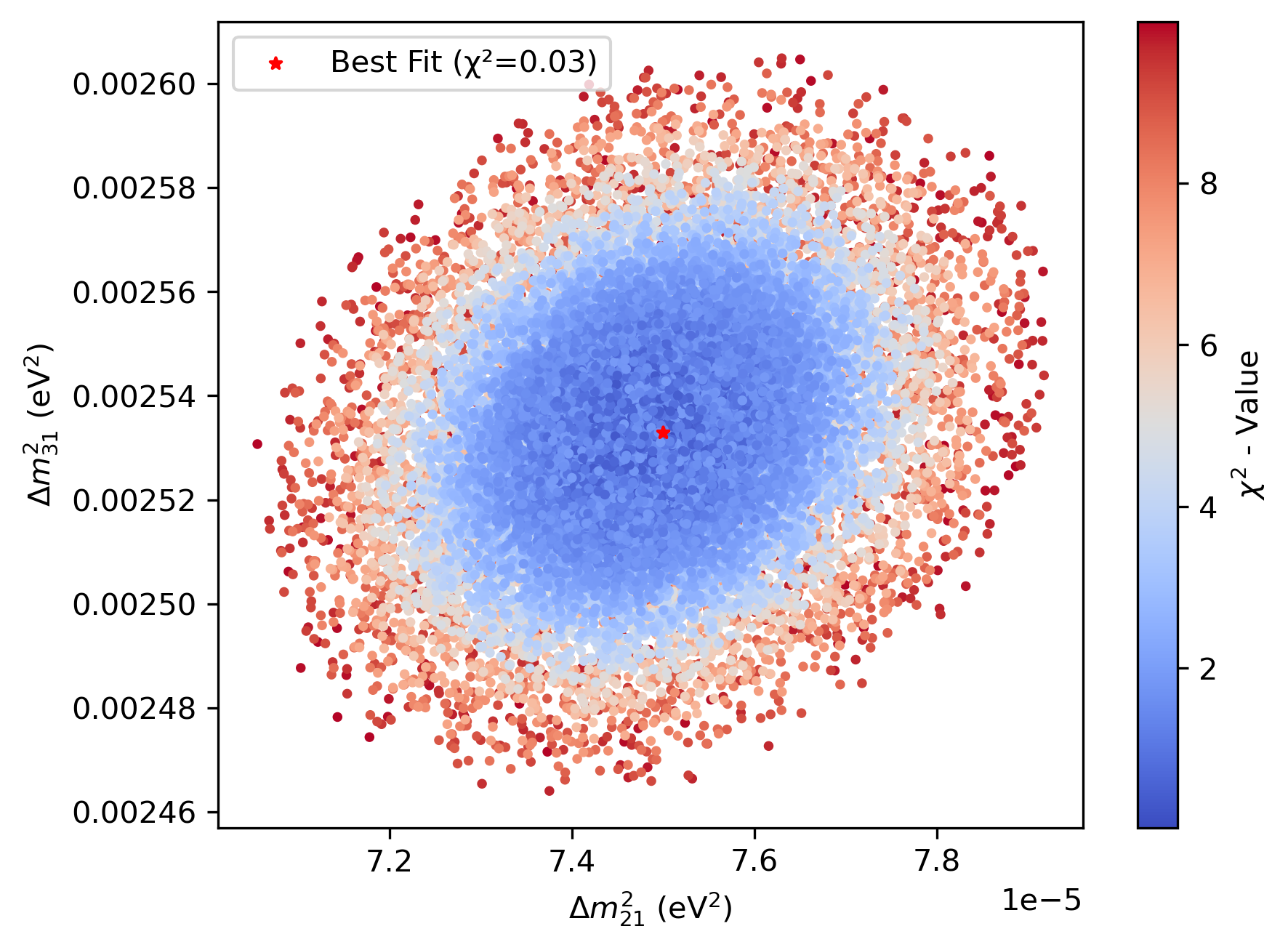}}
\quad
\subfigure[]{
\includegraphics[width=.45\textwidth]{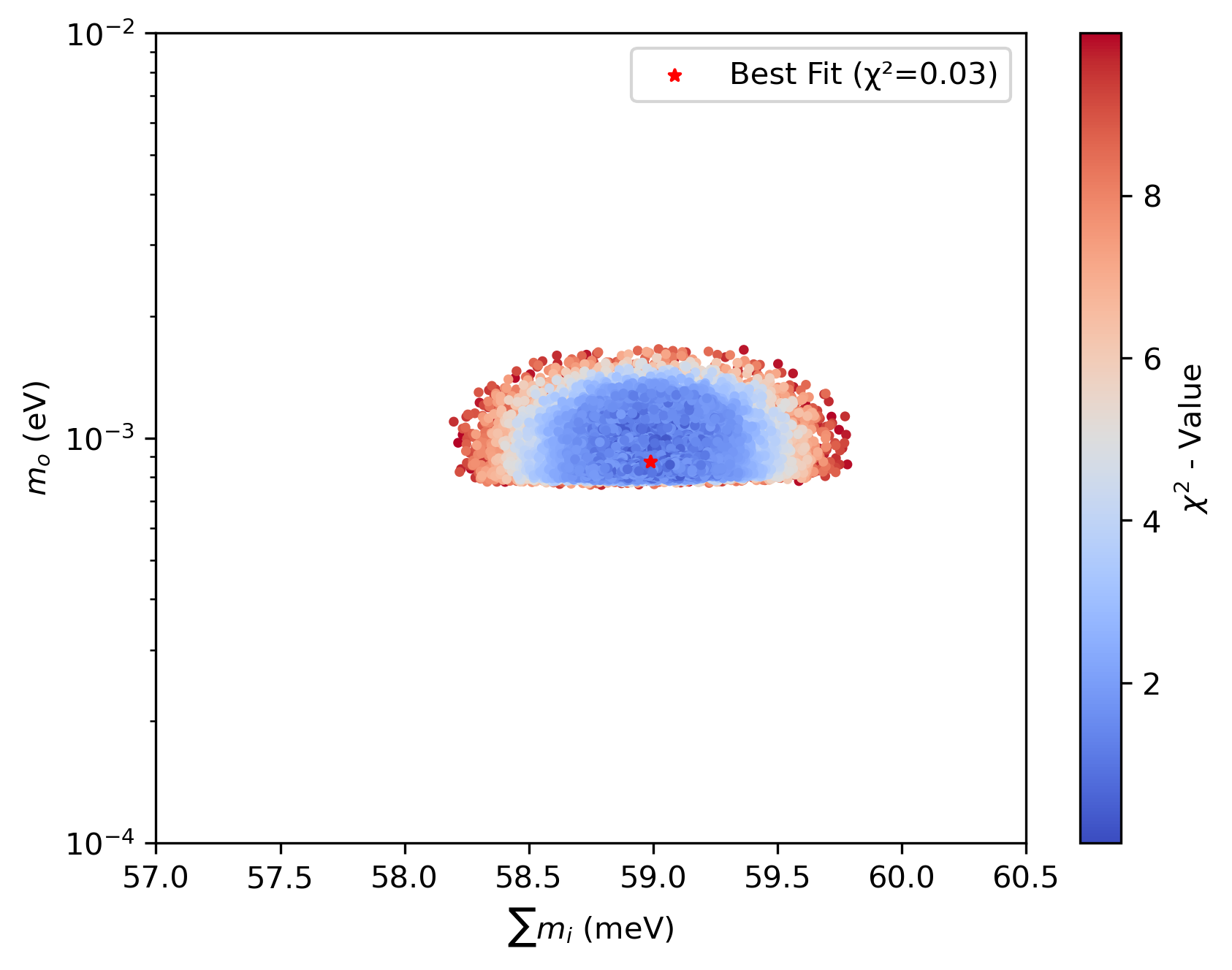}}
\quad
\vspace{-0.25 cm}
\caption{\footnotesize{Predicted values of neutrino mixing angles ($\sin^2\theta_{23},\sin^2\theta_{12}$, $\sin^2\theta_{13}$), mass squared differences $(\Delta m_{21}^2, \Delta m_{31}^2)$ and $m_{\Revised{0}}$.}}
\label{NOparam}
\end{figure}

The predictions of the model are calculated using the relations given in Eqs.(\ref{abcdgf}), (\ref{Neutrino_parameter}) and (\ref{Nij}). In the analysis, we have found that the considered model gives a good description of  of the neutrino oscillation data for NH with the best fit 
values of the model parameters correspond to a minimum value of $\chi^2$ with $\chi^2_{\mathrm{min}} = 0.03$.
In the case of IH, we observe a large best-fit $\chi^2_{min} > 100$ which predicts neutrino oscillation observables outside the experimental 3$\sigma $ range, i.e., the IH is not allowed in our model\footnote{Hereafter, the analysis is performed only for NH.}. The regions of free parameters allowed by the model along with their variations in $\chi^2$ values are shown in Figure \ref{abplot}. In these plots, the star symbol $\star$ (in bright red) represents the best fit point in each case. The best-fit values of the model parameters occur at

\bea
\mathrm{Re}\alpha =4.435,\,  \mathrm{Im}\alpha = -0.689,\, \mathrm{Re}\beta = 3.552,\, \mathrm{Im}\beta =-0.114,\, m_0 = 0.873\, \mathrm{meV}. \eea

\begin{figure}
\subfigure[]{
\includegraphics[width=.45\textwidth]{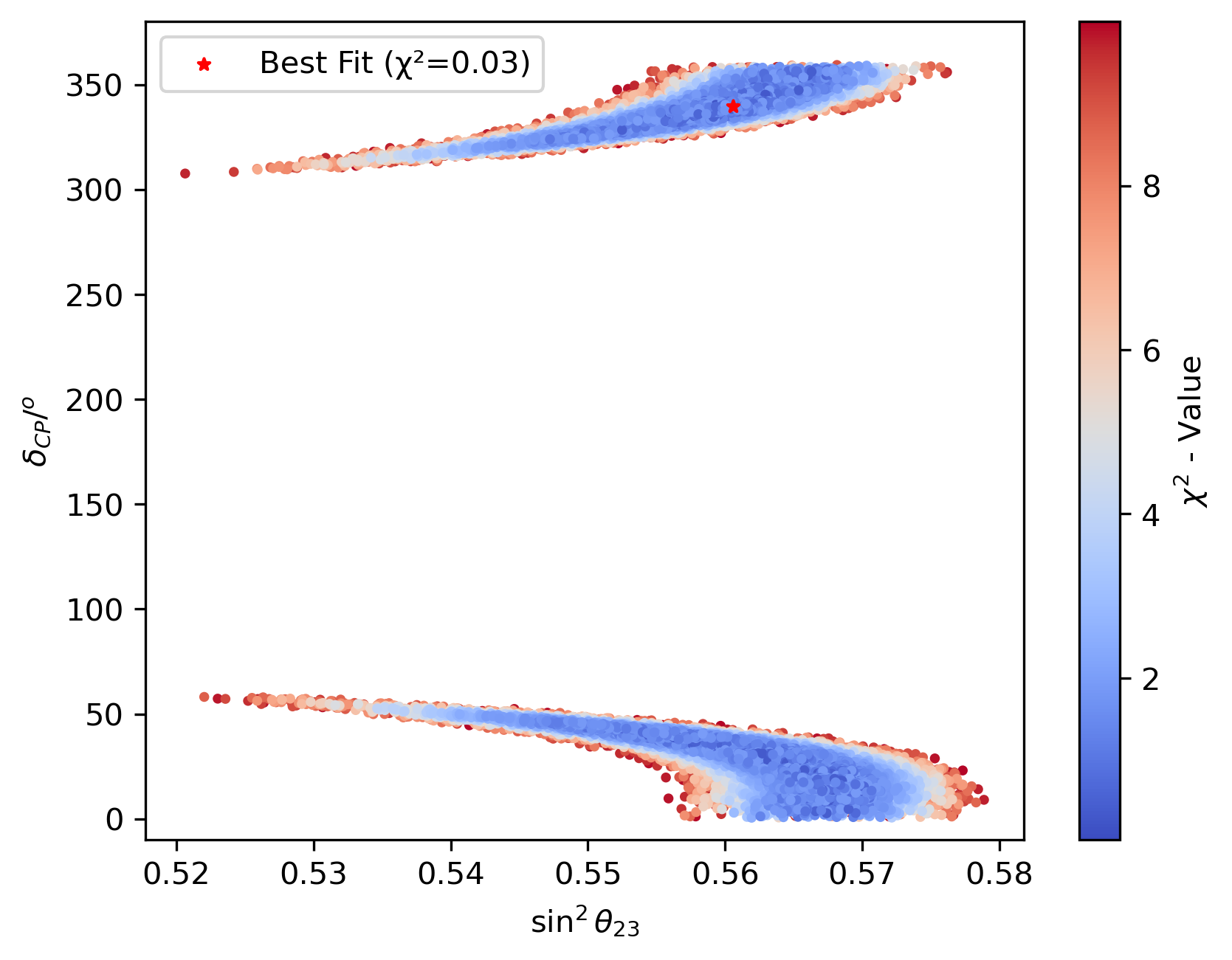}}
\quad
\subfigure[]{
\includegraphics[width=.45\textwidth]{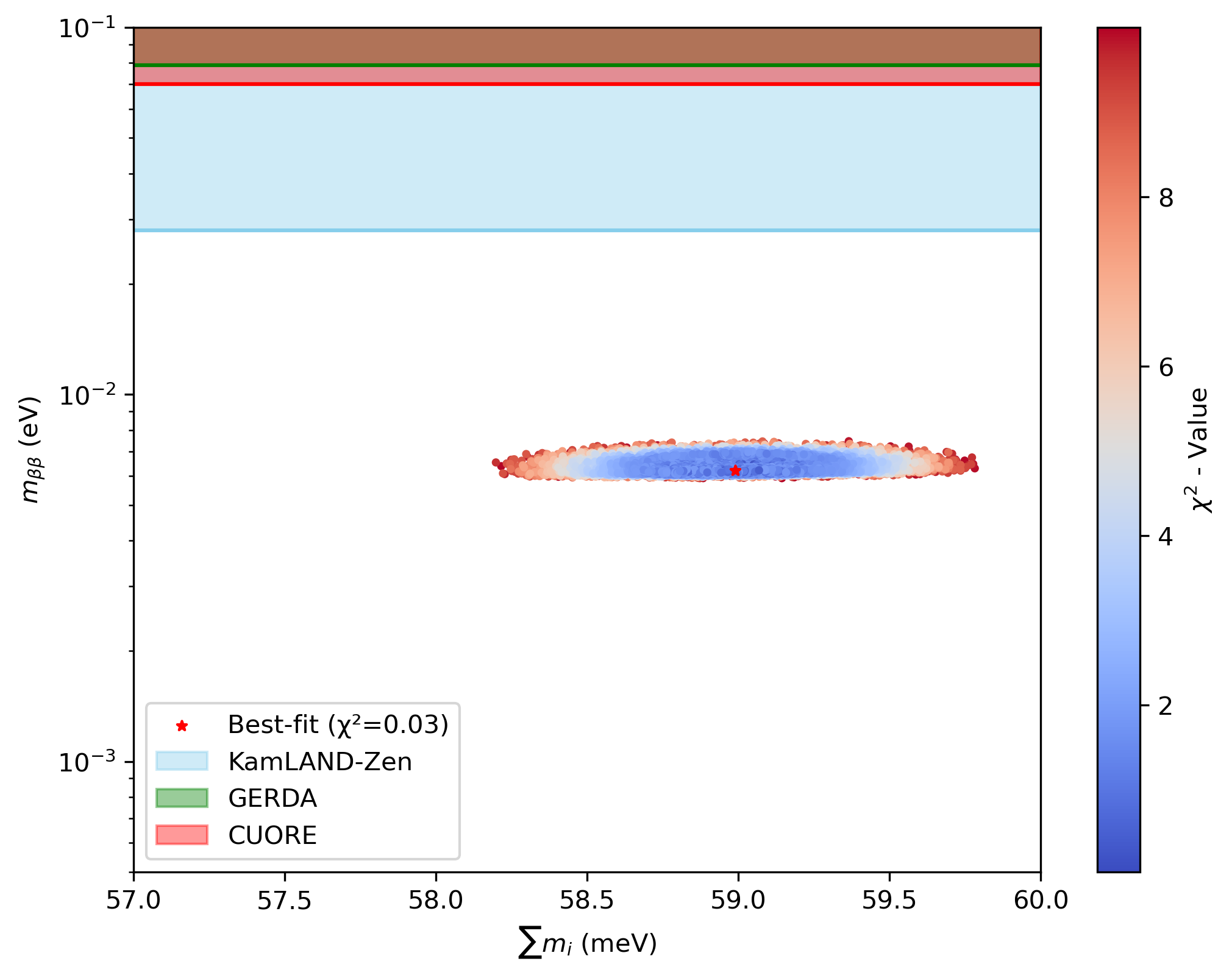}}
\quad
\vspace{-0.15 cm}
\caption{\footnotesize{Predicted values of Dirac CP-violating phase $\delta_{CP}$ and effective mass parameter $m_{\beta\beta}$. }}
\label{deltambb}
\end{figure}

The predictions of neutrino oscillation parameters are also shown as scatter plots in Figure \ref{NOparam}. We observe that the best-fit value of $\sin^2\theta_{23} = 0.560\, \, (\theta_{23}=48.40^\circ)$, suggesting a higher octant of $\theta_{23}$. 
The analysis  gives the sum of neutrino masses at 
$\sum = 58.98$ meV, which is also  consistent with the latest Planck Cosmological upper bound $\sum m_{\nu} < 0.072$ eV
\cite{DESI}. On the other hand, the model also predicts a Dirac CP-violating phase in two separate ranges, $\delta_{CP}\in (0.54,58.13)^\circ$ and $\delta_{CP}\in (307.61,359.35)^\circ$ with the best-fit value is $\delta_{CP}\simeq 339.81^\circ$, suggesting a lower half-plane of Dirac CP violation phase, as shown  in Figure \ref{deltambb}(a). The best-fit values of neutrino observables predicted by the model are summarised in Table \ref{bestfit}.
Furthermore, the model prediction of effective mass parameter $m_{\beta\beta}$ in the neutrinoless double beta decay is shown  in Figure \ref{deltambb}(b). The model predicts $m_{\beta\beta}$ in the range $(5.92 - 7.46)$ meV with the best-fit value is predicted at $m_{\beta\beta}^{\mathrm{bf}} = 6.20 $ meV.  These predictions are allowed by  the exclusion regions given by many experiments such as KamLAND-Zen, GERDA, CUORE, etc. However, future sensitivities of nEXO, LEGEND-1000 and CUPID which aim at probing the range $m_{\beta\beta} \in (4.7, 21.0) \,\mathrm{meV}$
will have a chance to reach the model prediction. 

\begin{table}[ht]
\vspace{0.5 cm}
\caption{Best-fit values of neutrino oscillation parameters predicted by the model at the minimum value $\chi^2_{\mathrm{min}} = 0.03$.}

\begin{tabular}{@{}ccccccccccc@{}}
\toprule
Parameters & $\sin^2\theta_{23}$& $\sin^2\theta_{12}$& $\sin^2\theta_{13}$& $\Delta m_{21}^2$\,(meV$^2$)& $\Delta m_{31}^2$\,(meV$^2$) &$m_2$\,(meV)& $m_3$\,(meV) &$\delta_{CP}^{(\circ)}$ & $r$ 
& $\sum$ (meV) \\
\hline
Best-fit & 0.560 & 0.306 & 0.0219 & 74.9 & $2.53\times 10^3$& 8.66 & 50.32 & 339.81 & 0.172 & 58.98 \\ \toprule
\end{tabular} \label{bestfit}
\vspace{0.25 cm}
\end{table}

In the numerical analysis of resonant leptogenesis, we do not consider the effects of scattering process, spectator effects, thermal corrections, e.t.c. In the considered model, the heavy right-handed Majorana neutrino is scanned as a free parameter in the range $a_R = [1 , 10]$ TeV 
while the light neutrino oscillation parameters are fixed at the best-fit values predicted by the model, given in Table \ref{bestfit}. The small Majorana mass term $a_{\mu}$ which is responsible for producing a non-zero splitting between the heavy masses is also 
considered as an 
input parameter. The other 
input parameters including Majorana phase $\sigma$, Re[$\zeta$] and Im[$\zeta$] are scanned in the following range,
\bea
\mathrm{Re}[\zeta] = [0 , 2\pi],\, \,  \mathrm{Im}[\zeta] = [-3 , 3],\, \, a_{\mu} = [10^{-6}, 10^{-2}]\, \mathrm{GeV}, \,\, a_R = [1 , 10] \, \mathrm{TeV},\, \, \sigma = [0, 2\pi]. \label{ReIm}\eea

\begin{figure}
\subfigure[]{
\hspace{-0.15 cm}\includegraphics[width=.435\textwidth]{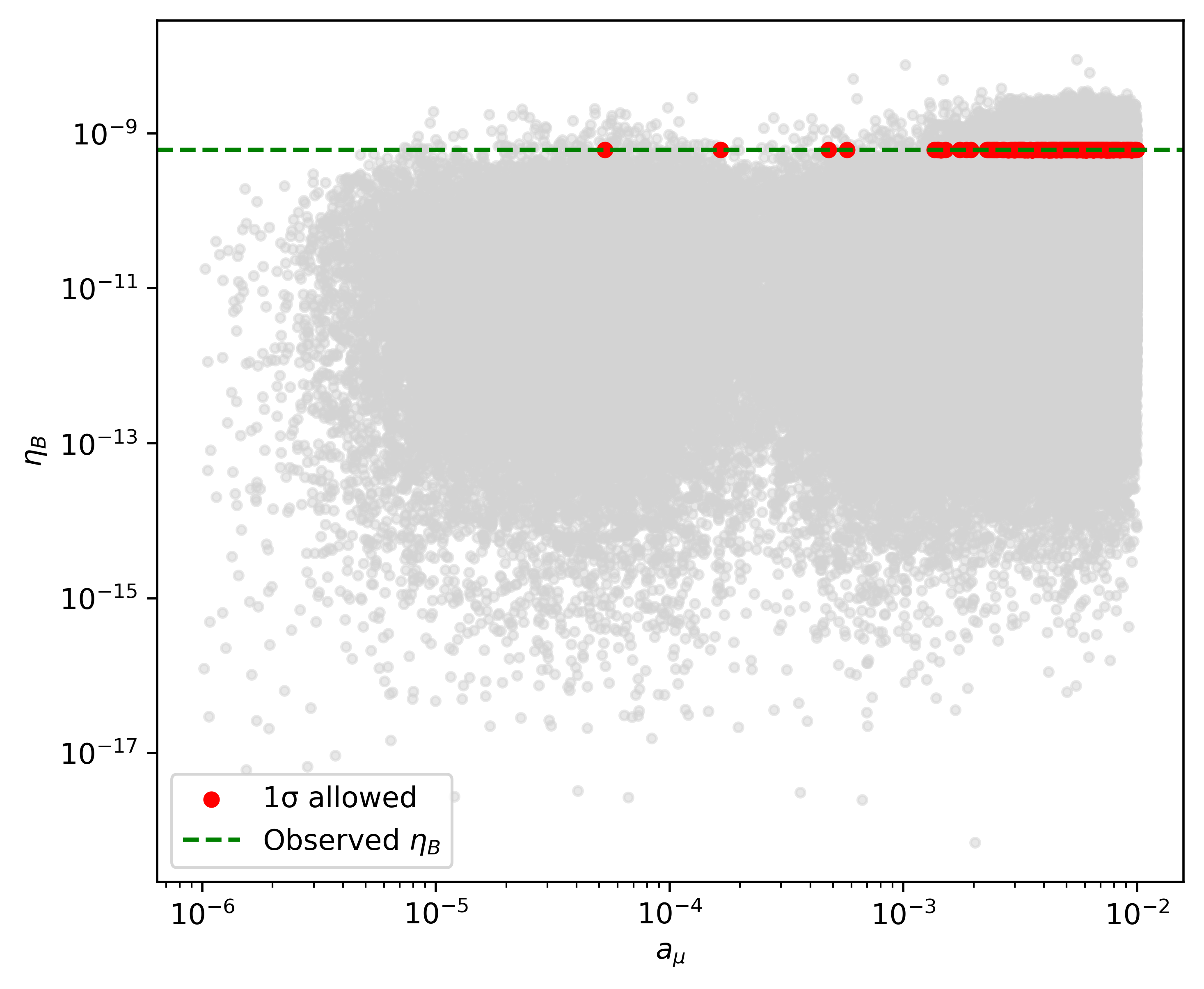}}
\quad
\subfigure[]{
\includegraphics[width=.45\textwidth]{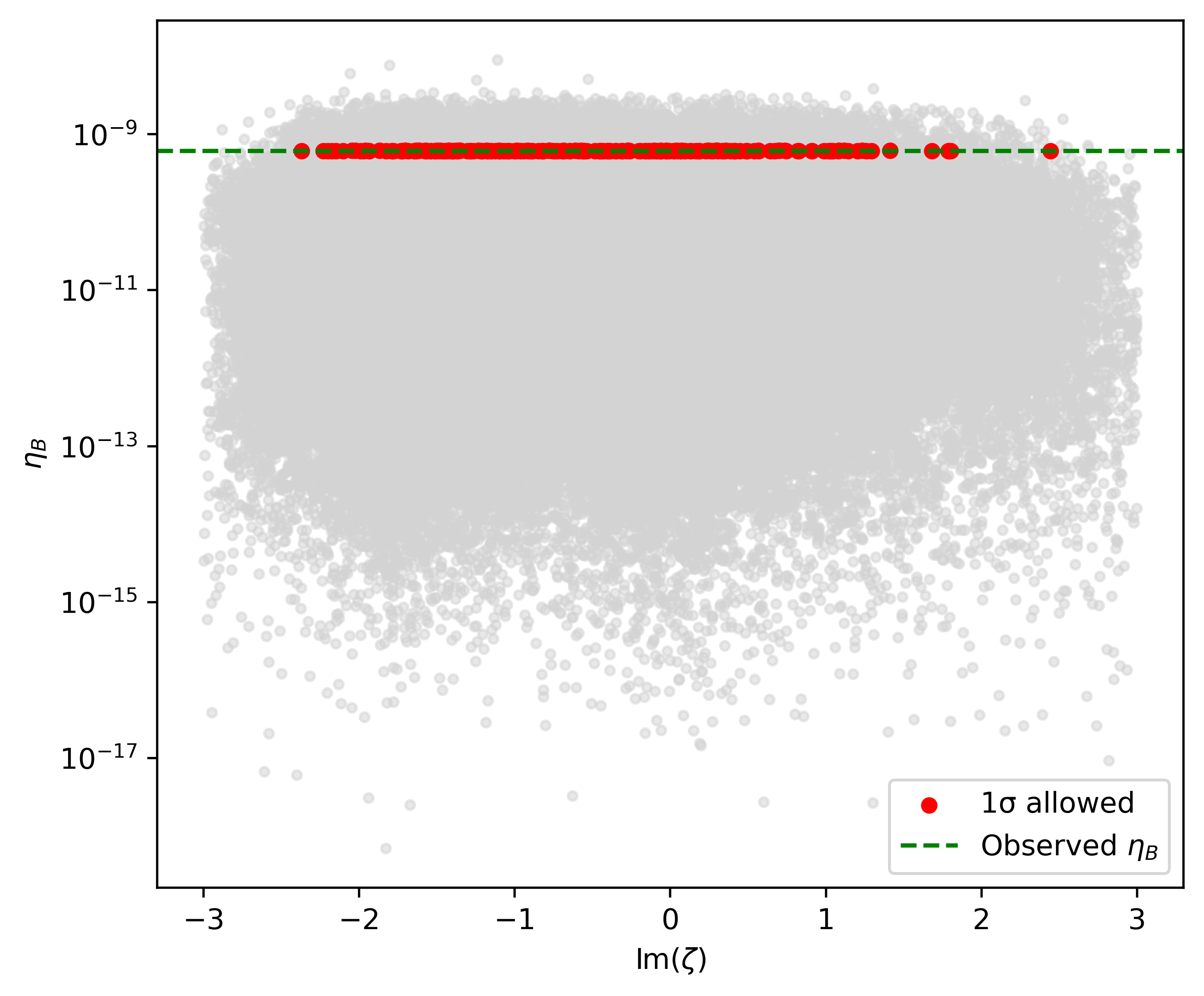}}
\quad
\subfigure[]{
\includegraphics[width=.45\textwidth]{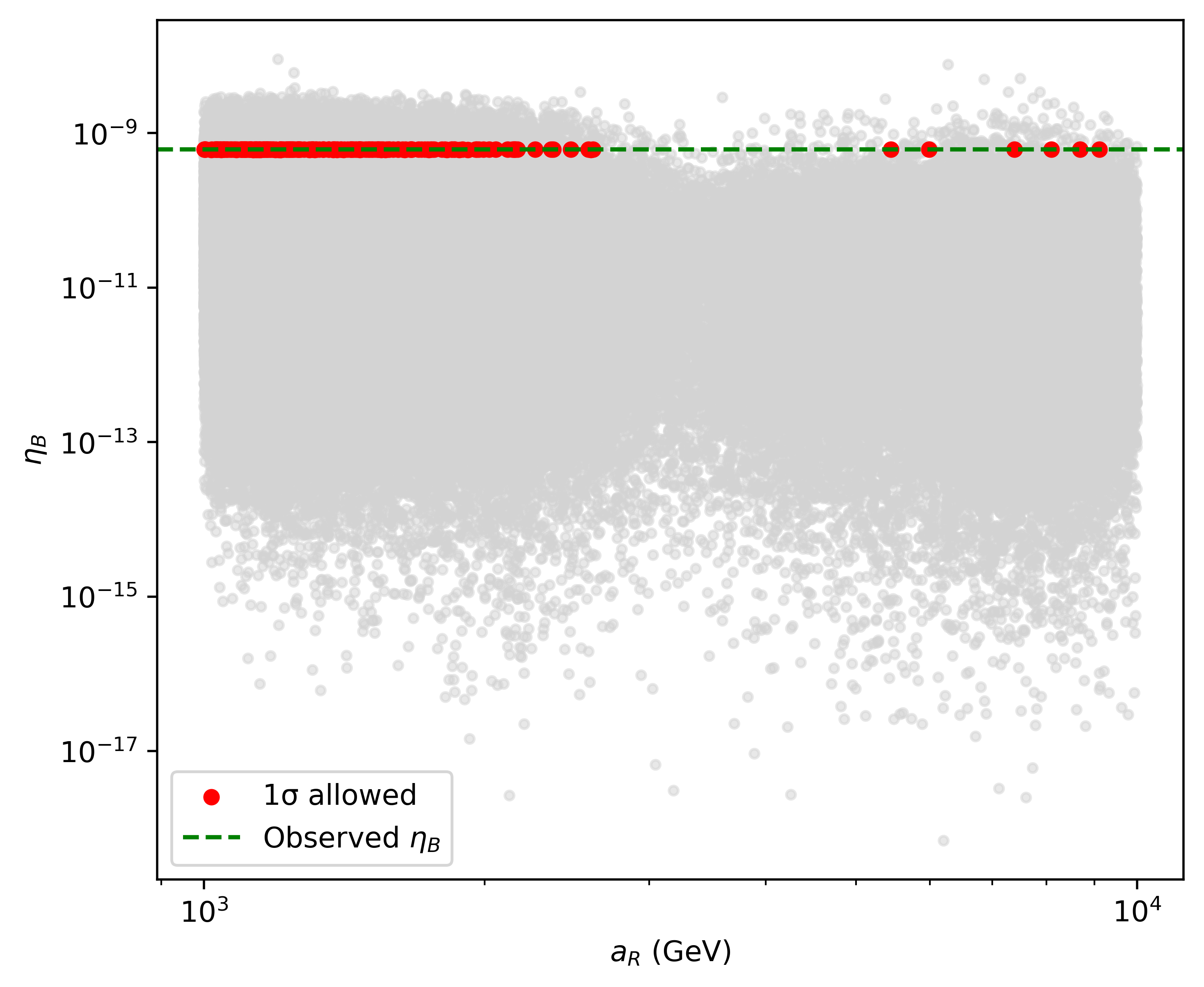}}
\quad
\subfigure[]{
\includegraphics[width=.45\textwidth]{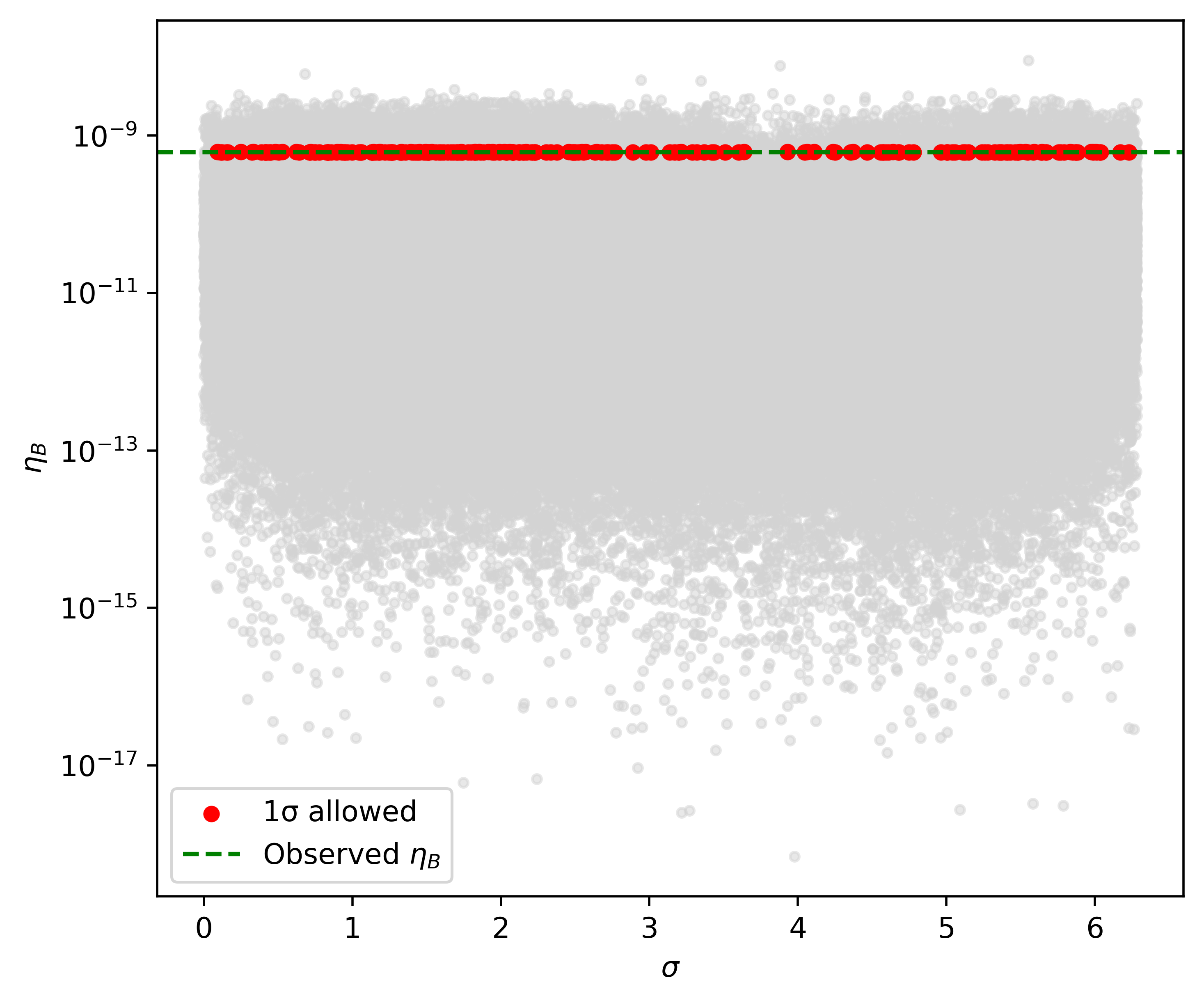}}
\quad
\vspace{-0.15 cm}
\caption{\footnotesize{Variation of $\eta_B$ with the model parameters. }}
\label{etaplot}
\end{figure}

\begin{figure}
\subfigure[]{
\includegraphics[width=.45\textwidth]{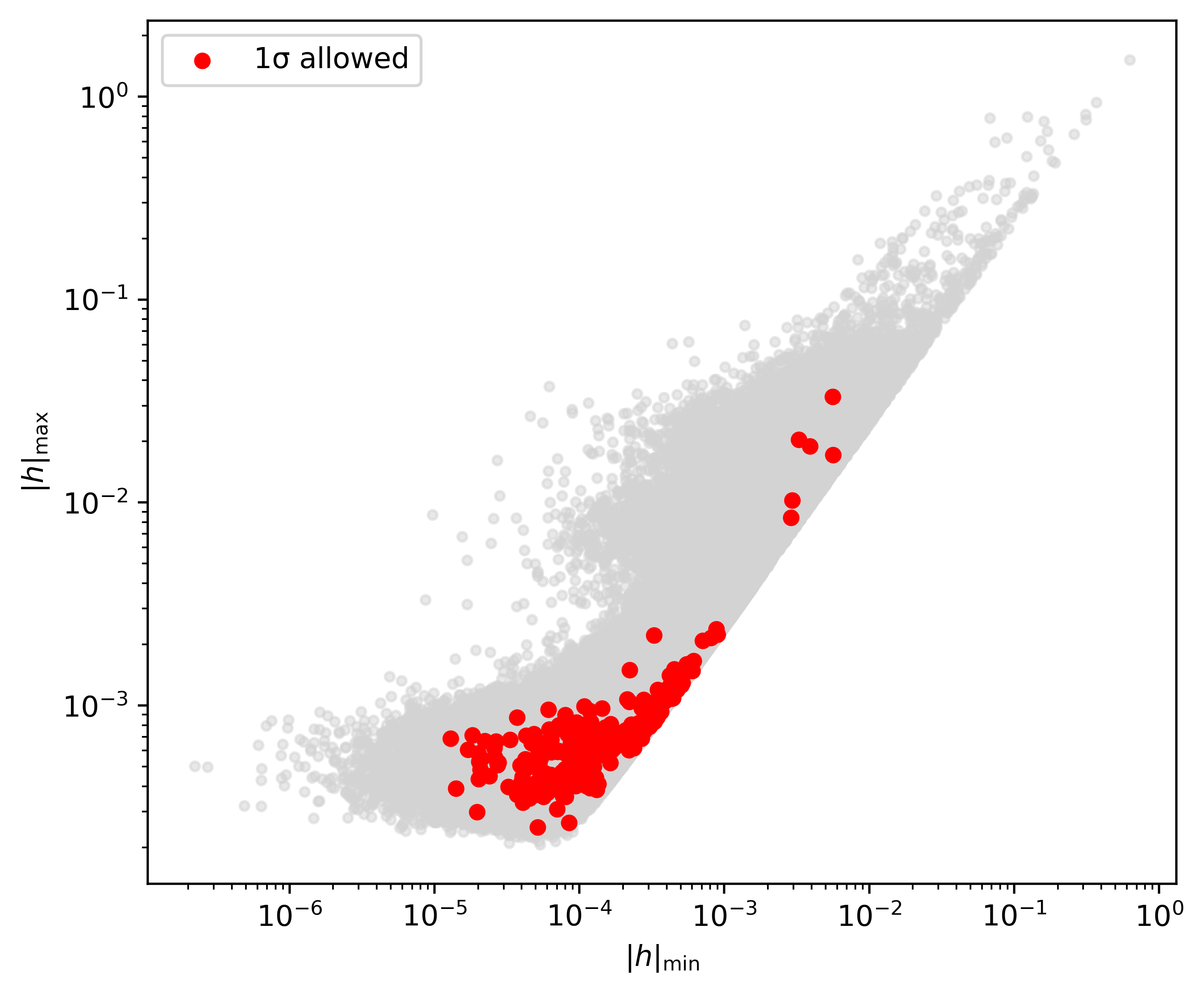}}
\quad
\subfigure[]{
\includegraphics[width=.45\textwidth]{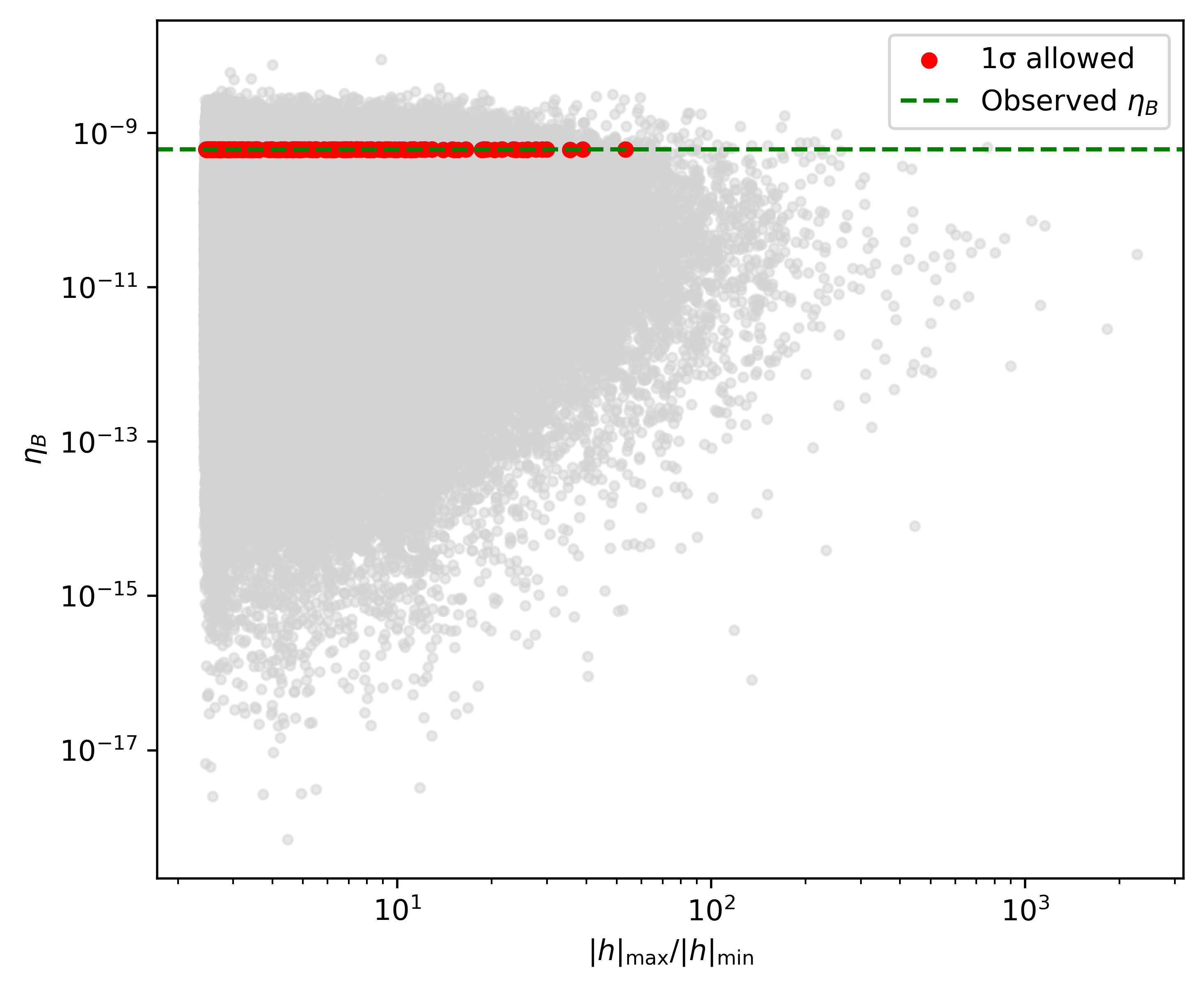}}
\quad
\vspace{-0.15 cm}
\caption{\footnotesize{Variation of Yukawa couplings. }}
\label{hplots}
\vspace{0.15 cm}
\end{figure}

\begin{figure}
\subfigure[]{
\hspace{-0.15 cm}\includegraphics[width=.435\textwidth]{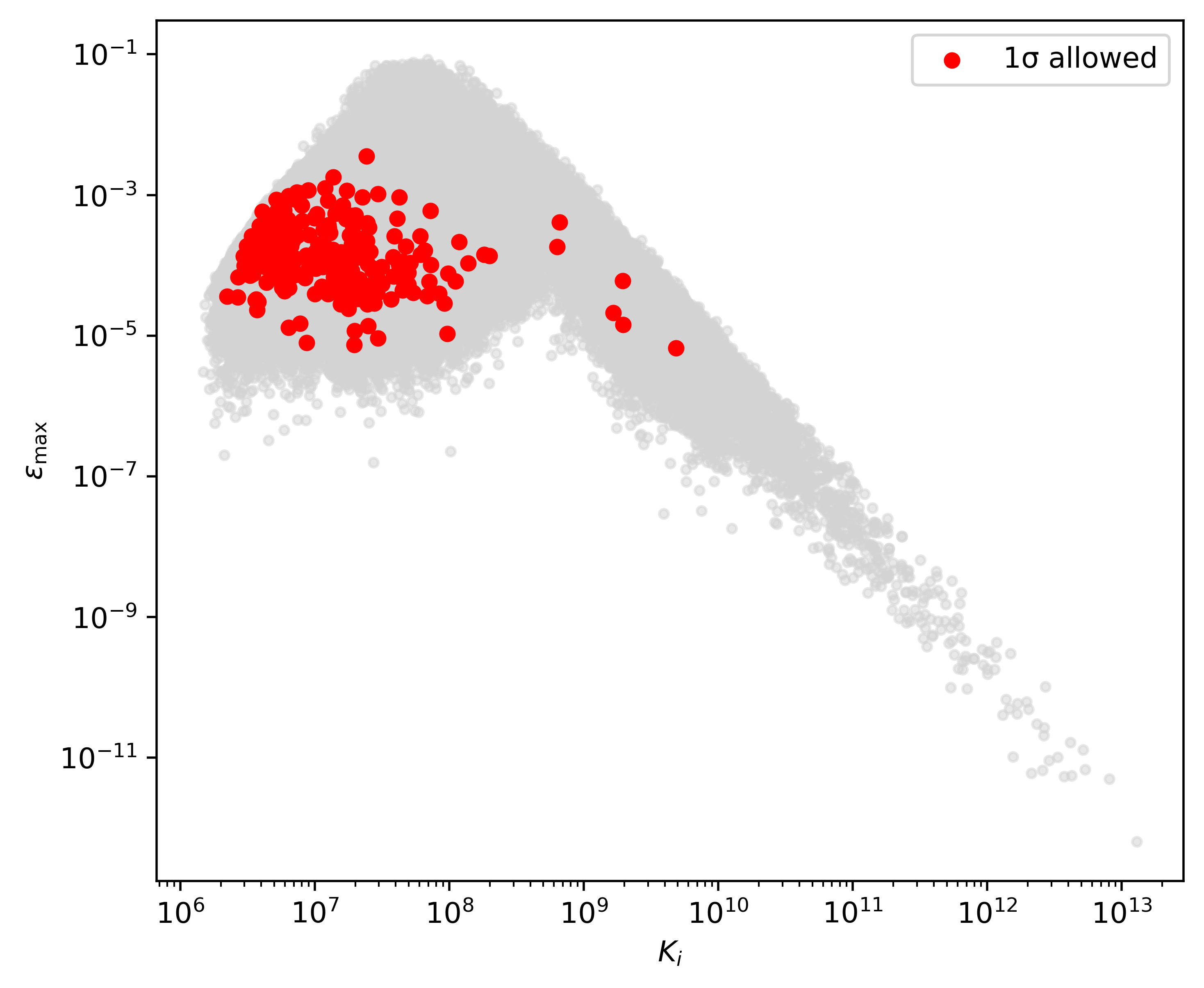}}
\quad
\subfigure[]{
\includegraphics[width=.45\textwidth]{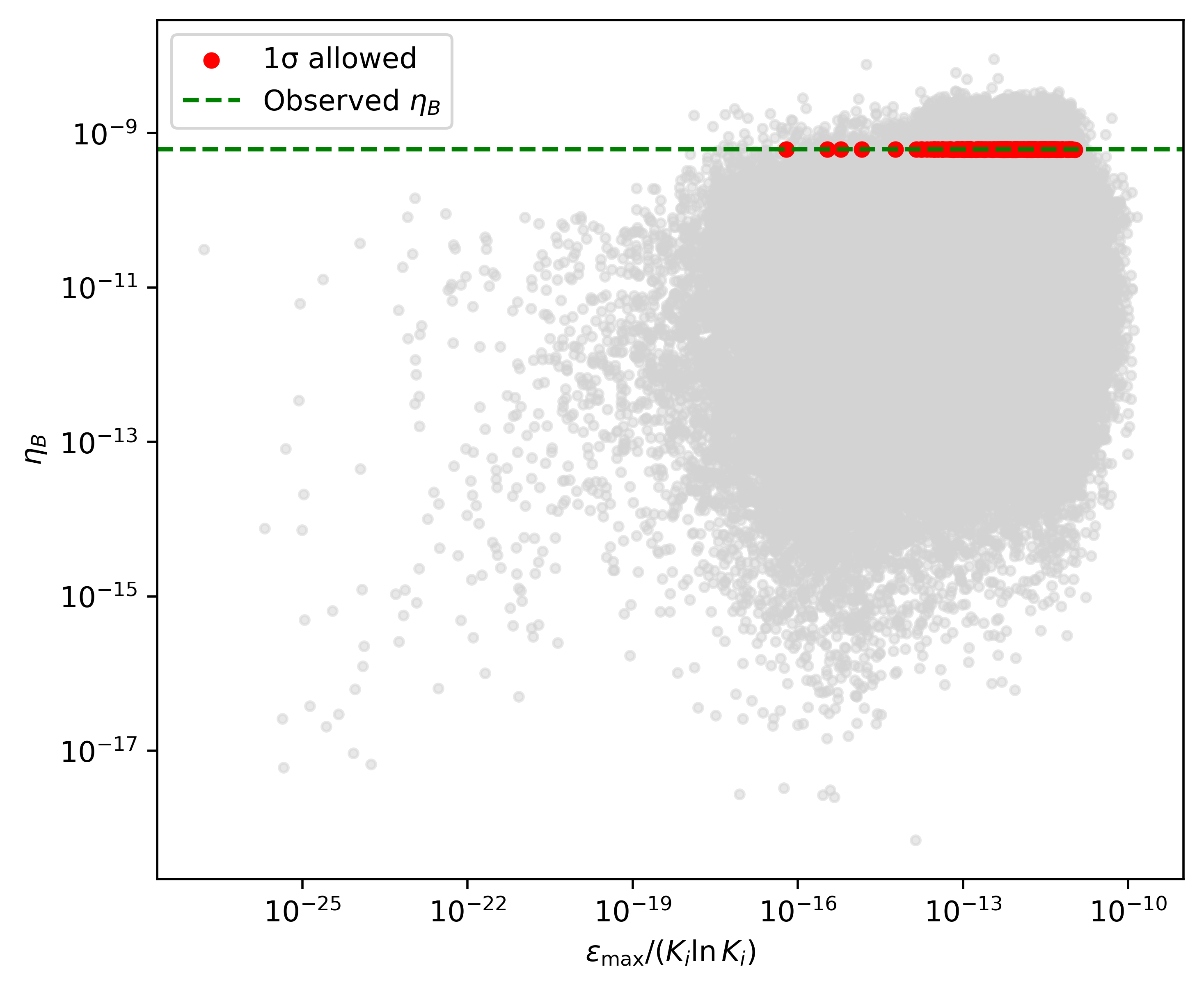}}
\quad
\subfigure[]{
\includegraphics[width=.45\textwidth]{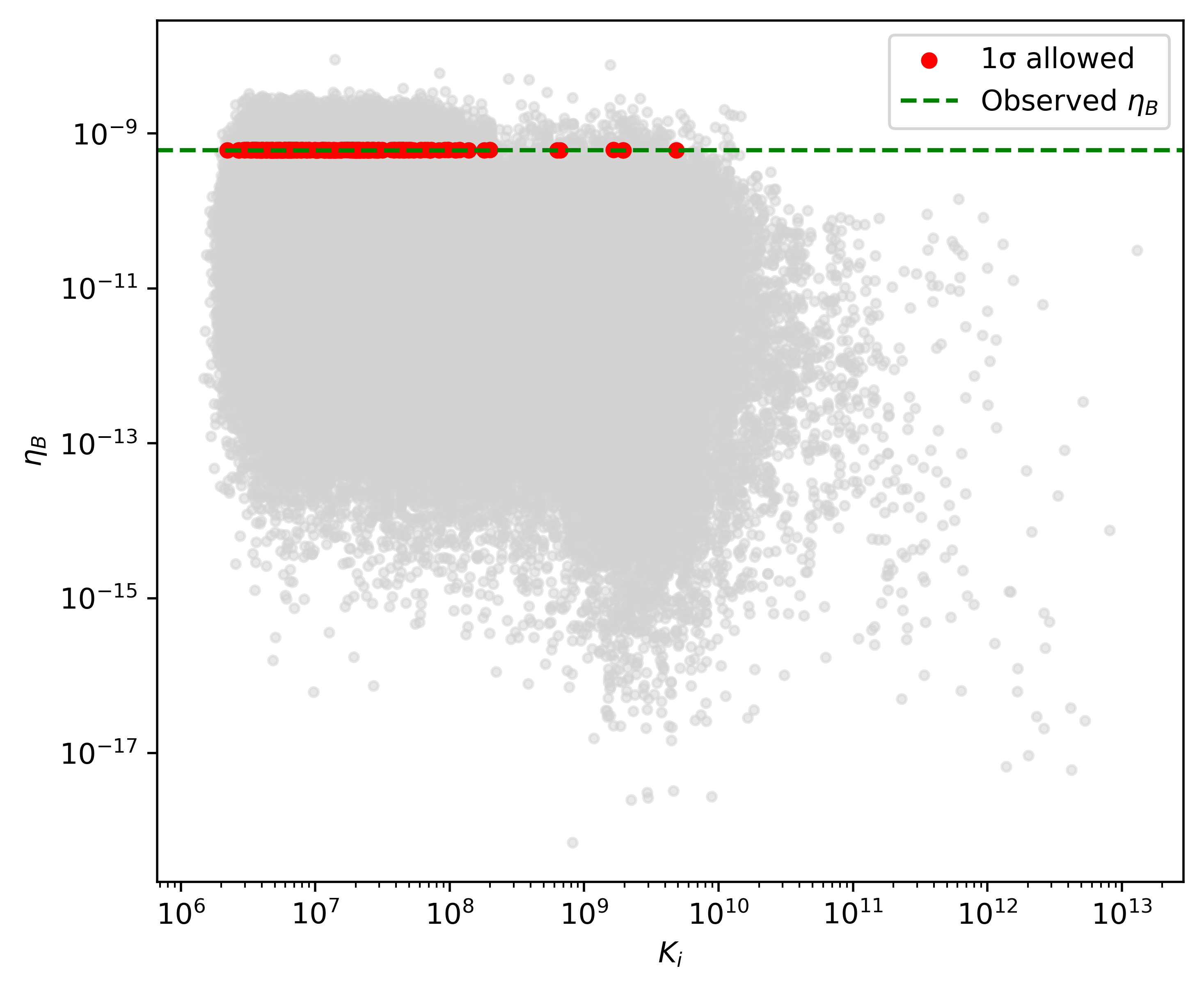}}
\quad
\subfigure[]{
\includegraphics[width=.45\textwidth]{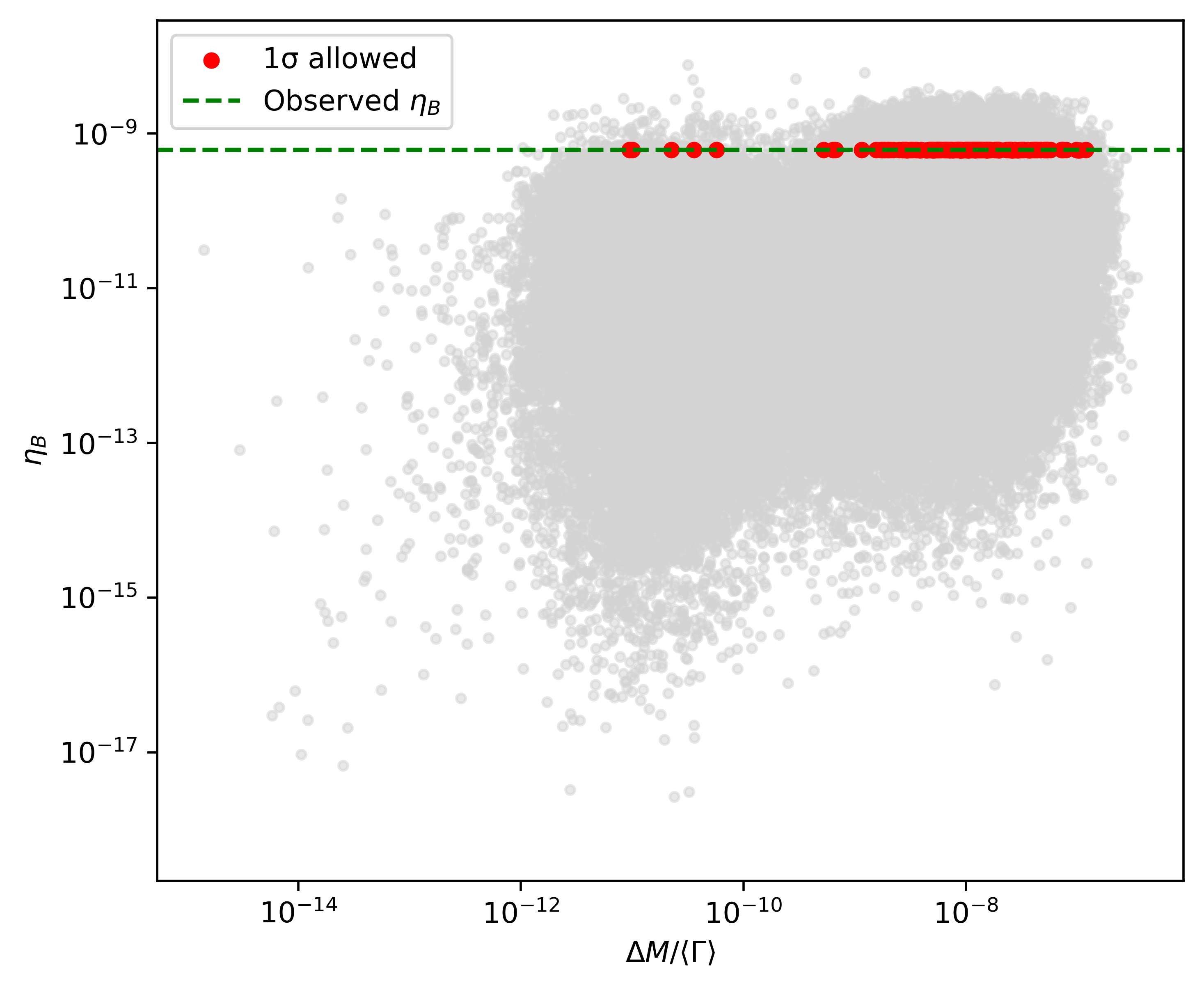}}
\quad
\vspace{-0.15 cm}
\caption{\footnotesize{Figure shows the variation between maximum CP asymmetry $\varepsilon_{\mathrm{max}}$} and effective washout factor $K_{i}$} in (a), $\eta_B$ with the ratio of maximum CP asymmetry and washout, $K_{i}$ and resonance ratio $\Delta M/ \langle \Gamma\rangle \sim a_{\mu}/\Gamma$.
\label{kplots}
\end{figure}

To explore a viable parameter space for successful resonant leptogenesis, we perform a Bayesian parameter scan using \textbf{Multinest} sampling package with a log-likelihood function defined as

\bea
\log L = -\frac{1}{2}\left( \frac{(\eta_B^i)^2 - (\eta_B^{obs})^2}{(\Delta \eta_B^{obs})^2}\right),
\eea
where, $\eta_B^i$ is the model predicted baryon asymmetry, $\eta_B^{obs} = 6.12\times 10^{-10}$ is the observed BAU 
 and  $\Delta \eta_B^{obs} = 0.04\times 10^{-10}$ is the 1$\sigma$ uncertainty. The results are presented in Figure \ref{etaplot}. In this figure, 
the highlighted red points correspond to the values of the model parameters that predict $\eta_B$ within the 1$\sigma$ range $(6.08,\,6.16)\times 10^{-10}$ and simultaneously satisfy the upper bounds on the cLFV processes discussed below. From Figure \ref{etaplot} we observe that successful leptogenesis is obtained for a narrow range of $a_{\mu}\sim (10^{-3}-10^{-2})$ GeV, while the heavy neutrino mass is preferred in the lower regions of $a_R \sim (1-3)$ TeV. Other parameters such as $\sigma$ and $\zeta$ do not have significant constraints on leptogenesis and a wide range of these parameters are allowed. The corresponding Yukawa couplings are observed within the perturbativity limits in the range $|h_{\alpha i}|\equiv \vert h\vert \sim 10^{-5}-10^{-2}$, as shown in Figure \ref{hplots}(a). We can also observe from Figure \ref{hplots}(b)
that the ratio of maximum and minimum Yukawa coupling  $\vert h_{\mathrm{max}}\vert / \vert h_{\mathrm{min}} \vert \sim (2.45 - 53.25)$ indicating comparable magnitudes of the Yukawa. Further, the maximum CP asymmetry is obtained in the range $\varepsilon_{\mathrm{max}} \sim 10^{-5} - 10^{-3}$ for successful leptogenesis. This result can be seen from Figure \ref{kplots}. From Figure \ref{kplots}(b), the ratio $\varepsilon_{\mathrm{max}}/(K_{i} \mathrm{ln} K_{i})$ is obtained in the range $10^{-14} - 10^{-11}$. In our analysis, the asymmetry is generated predominantly at late times, and the resonant enhancement of the CP asymmetry, supported by flavour projections in the range $P_{i\alpha} \sim 10^{-3} - 10^{-2},$ compensate the strong washout suppression and generate the observed value of $\eta_B$. However, from Figure \ref{kplots}(d), we observe that the maximal asymmetry does not coincide exactly with the general condition  $a_{\mu} = \Gamma$, since the final asymmetry is determined by the interplay between CP enhancement and washout effects within the full Boltzmann evolution in the analysis.

\begin{figure}
\subfigure[]{
\includegraphics[width=.45\textwidth]{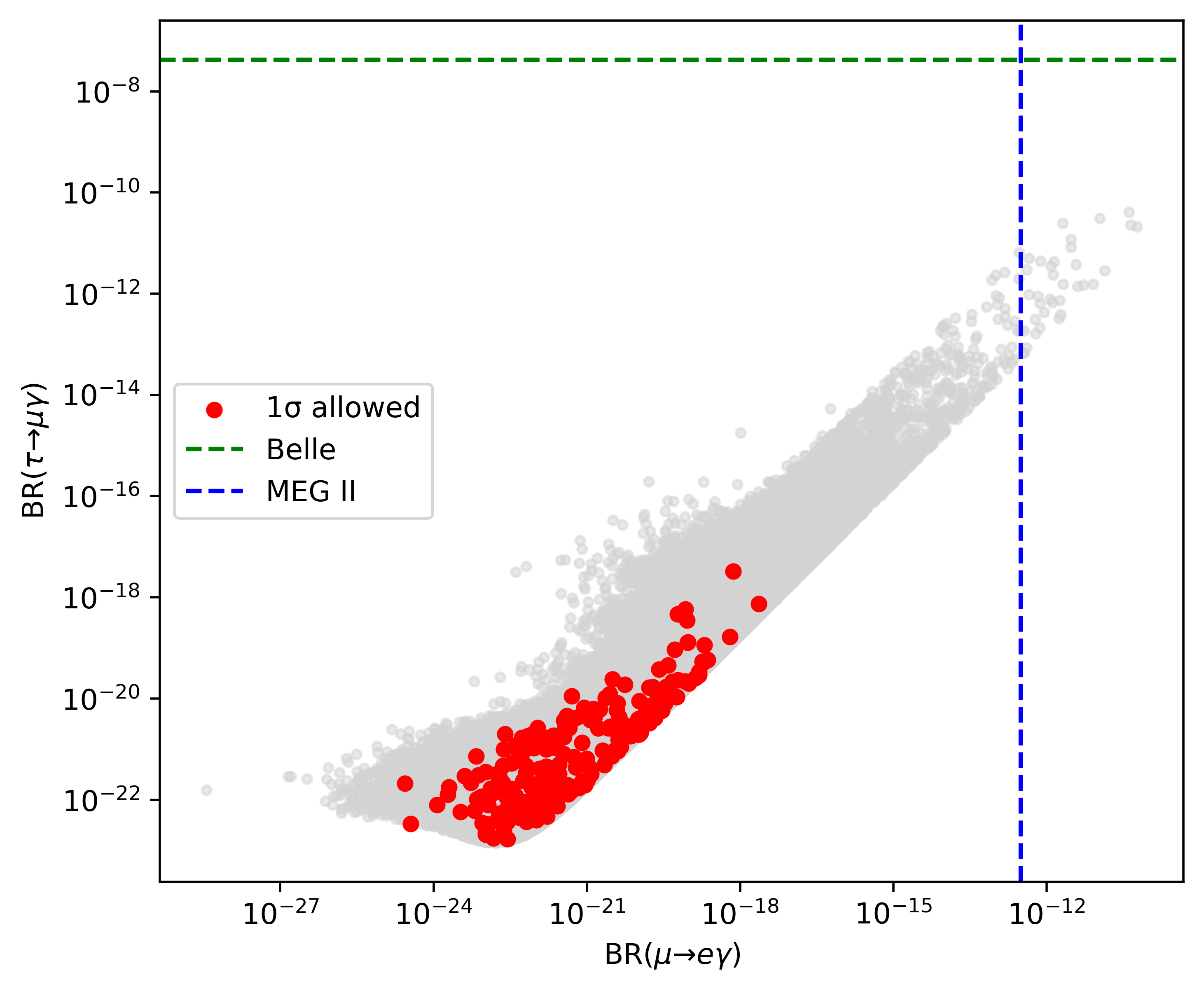}}
\quad
\subfigure[]{
\includegraphics[width=.45\textwidth]{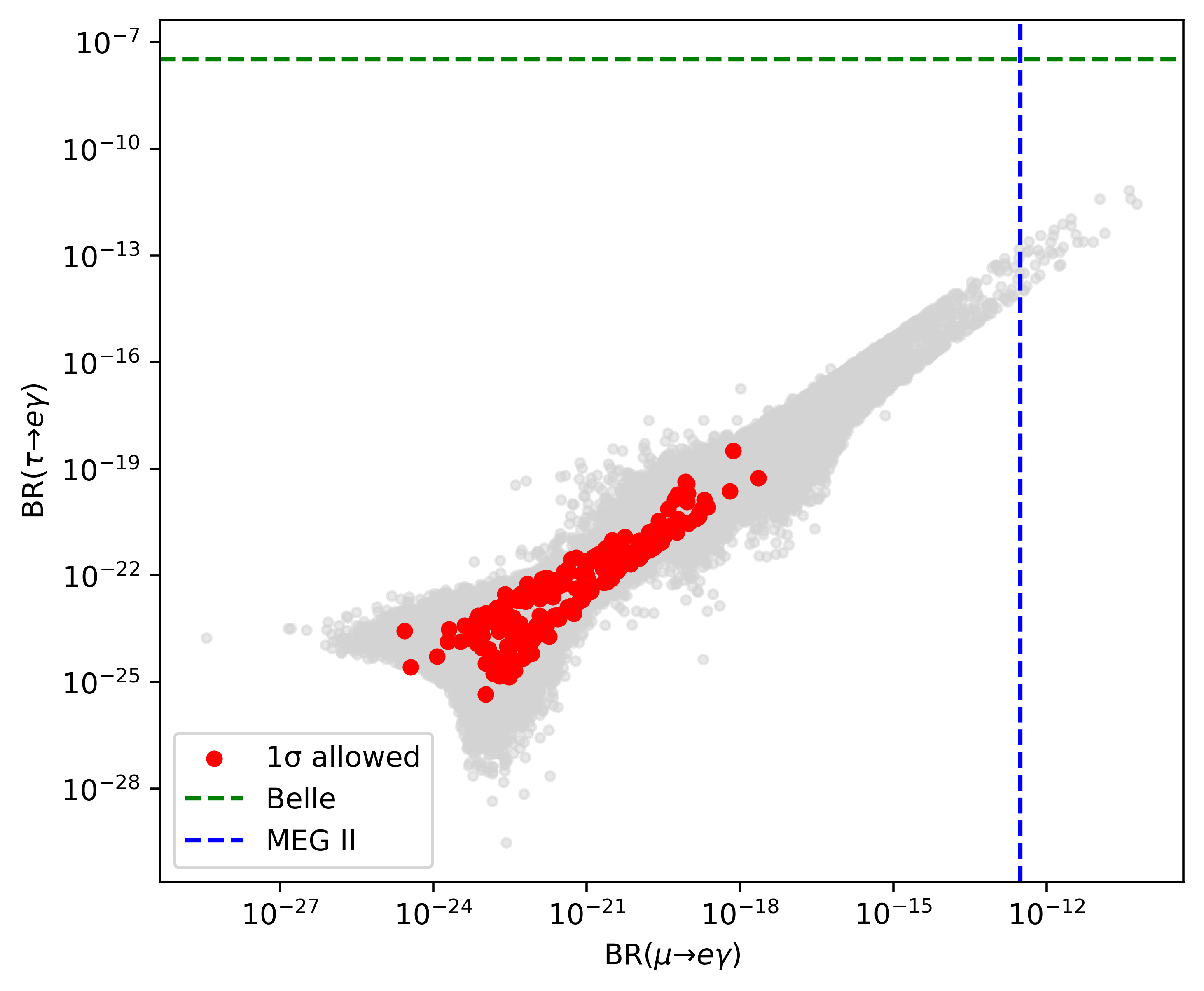}}
\quad
\vspace{-0.15 cm}
\caption{\footnotesize{Scatter plots between branching ratios in the study of cLFV processes. }}
\label{brplotfull}
\end{figure}

For the cLFV processes, the observed branching ratios of the model are shown in Figure \ref{brplotfull}. It is observed that the cLFV processes in our model are observed within the MEG II and Belle upper bounds. From the results shown in Figures \ref{etaplot}-\ref{kplots}, we can infer that the model successfully produces the observed BAU 
in the 1$\sigma$ range and also satisfies the MEG II, Belle and BaBar upper bounds on the cLFV decay processes for a wide range of the free parameters in Eq. (\ref{ReIm}).

\section{\label{conclusion} Conclusions}
We have proposed a minimal inverse seesaw model with $S_4$ symmetry for the Majorana neutrinos with only one real ($m_0$)-and two complex ($\alpha, \beta$) parameters in neutrino sector which gives reasonable predictions for the neutrino oscillation parameters, the observed baryon asymmetry of the Universe and the charged lepton flavor violation. The resulting model reveals a favor for normal mass ordering, a higher octant of $\theta_{23}$ with $s^2_{23}\simeq 0.560$ and a lower half-plane of Dirac CP violation phase with the best-fit value $\delta^{(\circ)}_{CP}\simeq 339.810$. The predictions of the model for sum of neutrino masses and the effective Majorana neutrino mass are centered around 58.98 meV and 6.2 meV, respectively. The future neutrino experiments such as T2K and NO$\nu$A will establish the octant of $\theta_{23}$ and provide a more precise measurement of Dirac CP-violation phase which can further strengthen the predictions of the model. The obtained masses of the heavy neutrinos at the MeV scale,  $M_{R}\sim 10^4\,$ GeV, could be testable by experiments in future. The model also provides the predictions of the baryon asymmetry and charged lepton flavour violation processes which are consistent with the experimental observations.

\section*{Acknowledgments}
This research is funded by Vietnam National Foundation for Science and Technology Development (NAFOSTED) under grant number 103.01-2023.45.
\newpage
\appendix
\section{\label{forbidappen}Forbidden Yukawa terms}
\begin{table}[h]
\begin{center}
\caption{Forbidden Yukawa terms}
\vspace{0.25 cm}
 \begin{tabular}{cccc} \toprule
 Yukawwa terms&Prevented by  \\ \hline
$(\overline{\psi}_L l_{1R})_{\underline{3}_1}H,
(\overline{\psi}_L l_{R})_{\underline{3}_1}(H\phi_l)_{\underline{3}_2}, (\overline{\psi}_L l_{R})_{\underline{3}_1} H, (\overline{\psi}_L l_{R})_{\underline{3}_2}H, (\overline{\psi}_L l_{R})_{\underline{3}_1}(H\phi_l)_{\underline{3}_2},$ &\multirow{3}{2 cm}{\hspace{0.8 cm}$S_4$}  \\
$(\overline{\psi}_L l_{R})_{\underline{3}_2}(H\varphi_l)_{\underline{3}_1},
(\overline{S} \nu_{R})_{\textbf{1}_1} (\varphi^{*}_l\chi)_{\textbf{3}_1},
(\overline{S} \nu_{R})_{\textbf{1}_2} (\varphi^{*}_l\chi)_{\textbf{3}_1},
(\overline{S} \nu_{R})_{\textbf{2}} (\varphi^{*}_l\chi)_{\textbf{3}_1},$&\\
$(\overline{S} \nu_{R})_{\textbf{1}_1} (\phi^{*}_l\chi)_{\textbf{3}_2},
(\overline{S} \nu_{R})_{\textbf{1}_2} (\phi^{*}_l\chi)_{\textbf{3}_2},
(\overline{S} \nu_{R})_{\textbf{2}} (\phi^{*}_l\chi)_{\textbf{3}_2},$& \\  \hline

$(\overline{\psi}_L l_{1R})_{\underline{3}_1}(H\varphi_\nu)_{\underline{3}_1},
(\overline{\psi}_L l_{1R})_{\underline{3}_1}(H\varphi^*_\nu)_{\underline{3}_1},
(\overline{\psi}_L l_{R})_{\underline{3}_1}(H\varphi_\nu)_{\underline{3}_1},
(\overline{\psi}_L l_{R})_{\underline{3}_1}(H\varphi^*_\nu)_{\underline{3}_1},$&\multirow{2}{2 cm}{\hspace{0.8 cm}$Z_5$}  \\
$(\overline{\psi}_L \nu_{R})_{\underline{3}_1}(\widetilde{H}\varphi_l)_{\underline{3}_1},
(\overline{\psi}_L \nu_{R})_{\underline{3}_2}(\widetilde{H}\phi_l)_{\underline{3}_2},
(\overline{\psi}_L \nu_{R})_{\underline{3}_1}(\widetilde{H}\varphi^*_\nu)_{\underline{3}_1},(\overline{\nu}^c_R \nu_{R})_{\textbf{1}_1}\chi^2,$&  \\ \hline
$(\overline{\nu}^c_R \nu_{R})_{\underline{1}_1}(\varphi^{*2}_\nu)_{\underline{1}_1},
(\overline{\nu}^c_R \nu_{R})_{\textbf{2}}(\varphi^{*2}_\nu)_{\textbf{2}},
(\overline{\nu}^c_R \nu_{R})_{\textbf{1}_1}(\varphi_l\varphi_\nu)_{\textbf{1}_1},
(\overline{\nu}^c_R \nu_{R})_{\textbf{2}}(\varphi_l\varphi_\nu)_{\textbf{2}},$&\multirow{2}{2 cm}{\hspace{0.8 cm}$Z_3$}  \\
$(\overline{\nu}^c_R \nu_{R})_{\textbf{1}_2}(\phi_l\varphi_\nu)_{\textbf{1}_2},
(\overline{\nu}^c_R \nu_{R})_{\textbf{2}}(\phi_l\varphi_\nu)_{\textbf{2}}$&\\ \hline
$(\overline{\psi}_L S^c)_{\underline{3}_1}(\widetilde{H}\varphi_l)_{\underline{3}_1},
(\overline{\psi}_L S^c)_{\underline{3}_2}(\widetilde{H}\phi_l)_{\underline{3}_2} $&$Z_2$  \\ \toprule 
\end{tabular}
\end{center}
\end{table}
\section{\label{Higgspotential} Scalar sector}
The total scalar potential, up to five dimensions, 
is given by\footnote{We use the notation $V(x_1 \rightarrow x_2, y_1 \rightarrow y_2) = V(x_1, y_1)\!\!\!\mid_{\{x_1=x_2,\, y_1=y_2\}}$.}:
\bea
V_{\mathrm{Scal}}&=& V(H) + V(\varphi_l)+ V(\phi_l)+ V(\varphi_\nu) + V(\chi)+V(H, \varphi_l)
+V(H, \phi_l)+V(H, \varphi_\nu)
+ V(H, \chi) \crn
&+&V(\varphi_l,\chi)+ V(\varphi_l, \phi_l)+V(\varphi_l, \varphi_\nu)
+ V(\phi_l, \varphi_\nu)+V(\phi_l,\chi) + V(\varphi_\nu, \chi) +V_{\mathrm{trip}}, \label{Vtotal}
\eea
where\footnote{$(\varphi_l^* \varphi_l)_{\textbf{2}}(\varphi_l^* \varphi_l)_{\textbf{2}}=0, (\varphi_l^* \varphi_l)_{\textbf{3}_{2a}}(\varphi_l^* \varphi_l)_{\textbf{3}_{2a}}=0, (\phi_l^* \phi_l)_{\textbf{2}}(\phi_l^* \phi_l)_{\textbf{2}}=0, (\phi_l^* \phi_l)_{\textbf{3}_{2a}}(\phi_l^* \phi_l)_{\textbf{3}_{2a}}=0, (\varphi_\nu^* \varphi_\nu)_{\textbf{3}_{2a}}(\varphi_\nu^* \varphi_\nu)_{\textbf{3}_{2a}}=0$, $(\varphi_l^* \varphi_l)_{\textbf{3}_{2a}}$ $(\phi_l^* \phi_l)_{\textbf{3}_{2a}}$=$(\varphi_l^* \varphi_l)_{\textbf{3}_{1s}} (\phi_l^* \phi_l)_{\textbf{3}_{2a}}=(\varphi_l^* \varphi_l)_{\textbf{3}_{2a}} (\phi_l^* \phi_l)_{\textbf{3}_{1s}}=0$, $(\varphi_l^* \varphi_l)_{\textbf{3}_{2a}} (\varphi_\nu^* \varphi_\nu)_{\textbf{3}_{2a}}=(\varphi_l^* \varphi_l)_{\textbf{3}_{1s}} (\varphi_\nu^* \varphi_\nu)_{\textbf{3}_{2a}}=(\varphi_l^* \varphi_l)_{\textbf{3}_{2a}} (\varphi_\nu^* \varphi_\nu)_{\textbf{3}_{1s}}=0$, $(\phi_l^* \phi_l)_{\textbf{3}_{1s}} (\varphi_\nu^* \varphi_\nu)_{\textbf{3}_{2a}}=(\phi_l^* \phi_l)_{\textbf{3}_{2a}} (\varphi_\nu^* \varphi_\nu)_{\textbf{3}_{1s}}=(\phi_l^* \phi_l)_{\textbf{3}_{2a}} (\varphi_\nu^* \varphi_\nu)_{\textbf{3}_{2a}}=0, (\varphi_l \phi_l)_\textbf{2}(\varphi_\nu^* \varphi_\nu)_\textbf{2}=0$ due to the VEV alignments of $\varphi_l, \phi_l$ and $\varphi_\nu$ and the antisymmetry of $\mathbf{3}_{2a}$ and $\mathbf{3}_{1a}$ as consequences of the tenser products of $\mathbf{3}_1\times \mathbf{3}_1, \mathbf{3}_2\times \mathbf{3}_2$ and $\mathbf{3}_1\times \mathbf{3}_2$ of $S_4$, respectively.}
\bea
&&V(H)=\mu^2_H H^\+H +\lambda^H (H^\+H)^2, 
V(\varphi_l)=\mu^2_{\varphi_l} (\varphi_l^* \varphi_l)_{\textbf{1}_1} +\lambda^{\varphi_l}_1 (\varphi_l^* \varphi_l)_{\textbf{1}_1} (\varphi_l^* \varphi_l)_{\textbf{1}_1}
+\lambda^{\varphi_l}_2 (\varphi_l^* \varphi_l)_{\textbf{3}_{1s}}(\varphi_l^* \varphi_l)_{\textbf{3}_{1s}}, \crn
&&V(\phi_l)=V(\varphi_l\rightarrow \phi_l), 
V(\varphi_\nu)=\mu^2_{\varphi_\nu} (\varphi_\nu^* \varphi_\nu)_{\textbf{1}_1} +\lambda^{\varphi_\nu}_1 (\varphi_\nu^* \varphi_\nu)_{\textbf{1}_1} (\varphi_\nu^* \varphi_\nu)_{\textbf{1}_1}
+\lambda^{\varphi_\nu}_2 (\varphi_\nu^* \varphi_\nu)_{\textbf{2}} (\varphi_\nu^* \varphi_\nu)_{\textbf{2}} \crn
&&\hspace{0.95 cm}+\lambda^{\varphi_\nu}_3 (\varphi_\nu^* \varphi_\nu)_{\textbf{3}_{1s}} (\varphi_\nu^* \varphi_\nu)_{\textbf{3}_{1s}}, \hs 
V(\chi)=\mu^2_\chi \chi^*\chi +\lambda^\chi (\chi^*\chi)^2, \hs 
V(H,\varphi_l)=\lambda^{H\varphi_l} (H^\+ H) (\varphi_l^* \varphi_l)_{\textbf{1}_1}, 
\crn
&&V(H,\phi_l)=V(H,\varphi_l\rightarrow \phi_l), \hs 
V(H,\varphi_\nu)=V(H,\varphi_l\rightarrow \varphi_\nu),\hs 
V(H,\chi)=\lambda^{H\chi} (H^\+ H) (\chi^* \chi), \crn
&&V(\varphi_l,\phi_l)=\lambda^{\varphi_l\phi_l}_1 (\varphi_l^* \varphi_l)_{\textbf{1}_1} (\phi_l^* \phi_l)_{\textbf{1}_1}
+\lambda^{\varphi_l\phi_l}_2 (\varphi_l^* \varphi_l)_{\textbf{3}_{1s}} (\phi_l^* \phi_l)_{\textbf{3}_{1s}}, 
V(\varphi_l,\varphi_\nu)=V(\varphi_l,\phi_l\rightarrow \varphi_\nu), 
\crn
&&
V(\varphi_l,\chi)=\lambda^{\varphi_l\chi} (\varphi_l^* \varphi_l)_{\textbf{1}_{1}} (\chi^* \chi), \hs
V(\phi_l,\varphi_\nu)=V(\varphi_l\rightarrow \phi_l,\varphi_\nu), \crn
&&V(\phi_l,\chi)=\lambda^{\phi_l\chi} (\phi_l^* \phi_l)_{\textbf{1}_{1}} (\chi^* \chi), \hs
V(\varphi_\nu,\chi)=V(\phi_l\rightarrow \varphi_\nu,\chi), 
\crn
&&V_{\mathrm{trip}}=\lambda^{H\varphi_l\varphi_\nu} (H^\dagger H) \big[\varphi_l(\varphi_\nu^* \varphi_\nu)_{3_{1s}}\big]_{\textbf{1}_{1}}
+\lambda^{\varphi_l\phi_l\varphi_\nu} (\varphi_l \phi_l)_{\textbf{3}_{2s}} (\varphi_\nu^* \varphi_\nu)_{\textbf{3}_{1s}}
+\lambda^{\varphi_l\varphi_\nu\chi} \big[\varphi_l(\varphi^*_\nu \phi_\nu)_{\textbf{3}_{1s}}\big]_{\textbf{1}_1} (\chi^* \chi). 
\eea
Now we will show that the VEVs in Eq. (\ref{VEV}) satisfy the
minimization condition of $V_{\mathrm{Scal}}$ by supposing that all the VEVs 
$\{v, v_{\varphi}, v_{\phi}, v_{1}, v_{2}, v_{3}, v_\chi\}\equiv v_\kappa$ are real. The minimum conditions of $V_{\mathrm{Scal}}$, $\frac{\partial V_{\mathrm{Scal}}}{\partial v_\kappa} =0$ and $\frac{\partial^2 V_{\mathrm{Scal}}}{\partial v_\kappa^2} >0$, yield the following relations: \\
\bea
&&\mu^2_H=-2 \lambda^{H} v^2-\lambda^{H\varphi_\nu} \left(v_{1}^2+2 v_{2} v_{3}\right)-\lambda^{H\varphi_l} v_{\varphi}^2-\lambda^{H\chi} v_{\chi}^2-\lambda^{H\phi_l} v_{\phi}^2 \crn
&&\hspace{0.55 cm}+\frac{2 (v_{1}^2-v_{2} v_{3}) \big[v_{\varphi} \big(2 \lambda_{2}^{\phi_l \varphi_\nu} v_{\varphi}+\lambda^{\varphi_l \varphi_\nu\chi} v_{\chi}^2\big)+2 \lambda_2^{\phi_l \varphi_\nu} v_{\phi}^2+2 \lambda^{\varphi_l \phi_l \varphi_\nu} v_{\varphi} v_{\phi}\big]}{v^2}, \label{muHsq}\\
&&\mu^2_{\varphi_l}=-\lambda^{H\varphi_l} v^2-v_{1}^2 (\lambda_1^{\varphi_l \varphi_\nu}+2 \lambda_2^{\phi_l \varphi\nu})
+2 v_{2}v_{3} (\lambda_2^{\phi_l\varphi_\nu}-\lambda_1^{\varphi_l \varphi_\nu})-2 v_{\varphi}^2 (\lambda_1^{\varphi_l}+4 \lambda_2^{\varphi_l})-\lambda^{ \varphi_l \chi}  v_{\chi}^2\crn
&&\hspace{0.6 cm}-v_{\phi}^2 (\lambda_1^{\varphi_l \phi}+4 \lambda_2^{\varphi_l \phi_l}) +\frac{2 \lambda_2^{\phi_l\varphi_\nu} v_{\phi}^2 \left(v_{1}^2-v_{2} v_{3}\right)}{v_{\varphi}^2}, \hspace{0.25 cm} \label{muvarlsq}\\
&&\mu^2_{\phi_l}=-\lambda^{H\phi_l} v^2-v_{1}^2 (\lambda_1^{\phi_l \varphi_\nu}+4 \lambda_2^{\phi_l \varphi_\nu})-2 \lambda_1^{\phi_l \varphi_\nu} v_{2} v_{3}+4 \lambda_2^{\phi_l \varphi_\nu} v_{2} v_{3}-v_{\varphi}^2 (\lambda_1^{\varphi_l \phi_l}+4 \lambda_2^{\varphi_l \phi_l})\crn
&&\hspace{0.6 cm}-\lambda^{\phi_l \chi}  v_{\chi}^2-2 v_{\phi}^2 (\lambda_1^{\phi}+4 \lambda_2^{\phi})-\frac{2 \lambda^{\varphi_l \phi_l \varphi_\nu} v_{\varphi} \left(v_{1}^2-v_{2} v_{3}\right)}{v_{\phi}}, \hspace{0.25 cm} \label{muphilsq}\\
&&\mu^2_{\varphi_\nu}=-\lambda^{H\varphi_\nu} v^2-2 (\lambda_1^{\varphi_\nu}+4 \lambda_3^{\varphi_\nu}) \left(v_{1}^2+2 v_{2} v_{3}\right)-\lambda_1^{\varphi_l \varphi_\nu} v_{\varphi}^2-\lambda^{\varphi_\nu \chi} v_{\chi}^2-\lambda_1^{\phi_l \varphi_\nu} v_{\phi}^2, \hspace{0.25 cm} \label{muvarlsq}\\
&&\mu^2_\chi=-v^2 \lambda^{H\chi}-\lambda^{\varphi_l \chi}  v_{\varphi}^2-2 \lambda^\chi  v_{\chi}^2-\lambda^{\phi_l \chi} v_{\phi}^2,\label{muchisq} \\
&&8 \lambda^{H} v^4+v_{1}^2 \left(8 \lambda_2^{\phi_l \varphi_\nu} v_{\varphi}^2+4 \lambda^{\varphi_l \varphi_\nu\chi} v_{\varphi} v_{\chi}^2+8 \lambda_2^{\phi_l \varphi_\nu} v_{\phi}^2+8 \lambda^{\varphi_l \phi \varphi_\nu} v_{\varphi} v_{\phi}-2 \lambda^{H \varphi_\nu} v^2\right)\crn
&&-4 v_{2} v_{3} \left(\lambda^{H\varphi_\nu} v^2+2 \lambda_2^{\phi_l \varphi_\nu} v_{\varphi}^2+\lambda^{\varphi_l \varphi_\nu \chi} v_{\varphi} v_{\chi}^2+2 \lambda_2^{\phi_l \varphi_\nu} v_{\phi}^2+2 \lambda^{\varphi_l \phi_l \varphi_\nu} v_{\varphi} v_{\phi}\right) >0, \label{ineq1}\\
&&-2 v_{1}^2 (\lambda_1^{\varphi_l \varphi_\nu}+2 \lambda_2^{\phi_l \varphi_\nu})+\frac{4 \lambda_2^{\phi_l \varphi_\nu} v_{\phi}^2 \left(v_{1}^2-v_{2} v_{3}\right)}{v_{\varphi}^2}+4 v_{2} v_{3} (\lambda_2^{\phi_l\varphi_\nu}-\lambda_1^{\varphi_l \varphi_\nu})+8 v_{\varphi}^2 (\lambda_1^{\varphi_l}+4 \lambda_2^{\varphi_l})>0, \label{ineq2}\\
&&-2 v_{1}^2 (\lambda_1^{\phi_l \varphi_\nu}+4 \lambda_2^{\phi_l \varphi_\nu})-\frac{4 \lambda^{\varphi_l \phi_l \varphi_\nu} v_{\varphi} \left(v_{1}^2-v_{2} v_{3}\right)}{v_{\phi}}-4 v_{2} v_{3} (\lambda_1^{\phi_l \varphi_\nu}-2 \lambda_2^{\phi_l \varphi_\nu})+8 v_{\phi}^2 (\lambda_1^{\phi_l}+4 \lambda_2^{\phi_l})>0. \label{ineq3}\\
&&\lambda_1^{\varphi_\nu}+4 \lambda_3^{\varphi_\nu}<0, \hs \lambda_{\chi}>0. \label{ineq4}
\eea

\end{document}